\documentclass[10pt]{article} 

\usepackage{latexsym,amssymb,amsfonts,amsmath,bbm} 

\usepackage{graphicx} 

\usepackage{cite} 

\usepackage{color} 

\usepackage[labelfont=bf,labelsep=period,justification=raggedright]{caption} 

\makeatletter 
\renewcommand{\@biblabel}[1]{\quad#1.} 
\makeatother

\date{} 

\graphicspath{{./}}

\renewcommand{\[}{\begin{equation}} 
\renewcommand{\]}{\end{equation}} 

\newcommand{\vect}[1]{\vec{#1}} 

\renewcommand{\O}{\operatorname{\mathbb{O}}} 
 
\newcommand{\E}{\operatorname{\mathbb{E}}} 
 
\newcommand{\Var}{\operatorname{\mathbb{V}\hspace{-0.6pt}\mathrm{ar}}}

\newcommand{\one}{\mathbbm{1}}

\renewcommand{\deg}{^{\circ}} 
\newcommand{\erf}{\operatorname{\mathrm{erf}}}

\usepackage{listings} 
\usepackage{setspace} 
\usepackage{textcomp} 

\begin{document} 

\begin{flushleft} 
{\Large 
\textbf{Distribution of Orientation Selectivity in Recurrent Networks of Spiking Neurons with Different Random Topologies} 
} 
\\ 
Sadra Sadeh$^{1}$, 
Stefan Rotter$^{1,\ast}$ 
\\ 
\bf{1} Bernstein Center Freiburg \& Faculty of Biology, University of Freiburg, Germany 
\\ 
$\ast$ E-mail: stefan.rotter@biologie.uni-freiburg.de 
\end{flushleft} 


\section*{Abstract} 

Neurons in the primary visual cortex are more or less selective for the orientation of a light bar used for stimulation. 
A broad distribution of individual grades of orientation selectivity has in fact been reported in all species. 
A possible reason for emergence of broad distributions is the recurrent network within which the stimulus is being processed. 
Here we compute the distribution of orientation selectivity in randomly connected model networks that are equipped with different spatial patterns of connectivity. 
We show that, for a wide variety of connectivity patterns, a linear theory based on firing rates accurately approximates the outcome of direct numerical simulations of networks of spiking neurons. 
Distance dependent connectivity in networks with a more biologically realistic structure does not compromise our linear analysis, as long as the linearized dynamics, and hence the uniform asynchronous irregular activity state, remain stable. 
We conclude that linear mechanisms of stimulus processing are indeed responsible for the emergence of orientation selectivity and its distribution in recurrent networks with functionally heterogeneous synaptic connectivity. 




\section*{Introduction} 
\label{Sec_Intro} 

When arriving at the cortex from the sensory periphery,  sensory signals are further processed by local recurrent networks. 
Indeed, the vast majority of all the connections a cortical neuron receives are from the cortical networks within which it is embedded and only a small fraction of connections are from feedforward afferents: The fraction of recurrent connections has been estimated to be as large as $80\%$ \cite{Peters1993}. 
What is the precise role of this recurrent network in sensory processing is not yet fully clear. 

In the primary visual cortex of mammals like carnivores and primates, for instance, it has been proposed that the recurrent network might be mainly responsible for the amplification of orientation selectivity \cite{Ben-Yishai1995, Somers1995}. 
Only a small bias provided by the feedforward afferents would be enough, and selectivity is then amplified by a non-linear mechanism implemented by the recurrent network. 
This mechanism is a result of the feature-specific connectivity assumed in the model, where neurons with similar input selectivities are connected to each other with a higher probability. This, in turn, could follow from the arrangement of neurons in orientation maps \cite{Blasdel1986, Bonhoeffer1991, Ohki2006}, which implies that nearby neurons have similar preferred orientations. As nearby neurons are also connected with a higher likelihood than distant neurons, feature-specific connectivity is a straight-forward result in this scenario. 

Feature-specific connectivity is not evident in all species, however. 
In rodent visual cortex, for instance, a salt-and-pepper organization of orientation selectivity has been reported, with no apparent spatial clustering of neurons according to their preferred orientations \cite{Ohki2006}. 
As a result, each neuron receives a heterogeneous input from pre-synaptic sources with different preferred orientations \cite{Jia2010}. 
 
Although an over-representation of connections between neurons of similar preferred orientations has been reported in rodents \cite{Hofer2011, Ko2011, ko2013, Ko2014, Ishikawa2014}, presumably as a result of a Hebbian growth process during a later stage of development \cite{Harris2013}, such feature-specific connectivity is not yet statistically significant immediately after eye opening \cite{ko2013}. 
A comparable level of orientation selectivity, however, has indeed been reported already at this stage \cite{ko2013}. If cortical recurrent networks make a contribution to sensory processing at this stage, random recurrent networks should be chosen as a model \cite{Hansel2012, Pehlevan2014, Sadeh2014}. Activity-dependent reorganization of the network, however, may still refine the connectivity and improve the performance of the processing later during development. 

Here we study the distribution of orientation selectivity in random recurrent networks with heterogeneous synaptic projections, i.e.\ networks where the recurrent connectivity does not depend on the preferred feature of the input to the neurons. 
We show that in structurally homogeneous networks, the heterogeneity in functional connectivity, i.e.\ the heterogeneity in preferred orientations of recurrently connected neurons, is indeed responsible for a broad distribution of selectivities. 
A linear analysis of the network operation can account quite precisely for this distribution, for a wide range of network topologies including Erd\H{o}s-R\'{e}nyi random networks and networks with distance-dependent connectivity. 


\section*{Methods}

\subsection*{Network Model} 

In this study, we consider networks of leaky integrate-and-fire (LIF) neurons. 
For this spiking neuron model, the sub-threshold dynamics of the membrane potential $V_i(t)$ of neuron $i$ is described by the leaky-integrator equation 
\[\tau \dot{V}_i(t) + V_i(t) = RI_i(t). 
  \label{Eq_NeuronModel} 
\] 
The current $I_i(t)$ represents the total input to the neuron, the integration of which is governed by the leak resistance $R$, and the membrane time constant $\tau = 20\,\mathrm{ms}$. 
When the voltage reaches the threshold at $V_\mathrm{th} = 20\,\mathrm{mV}$, a spike is generated and transmitted to all postsynaptic neurons, and the membrane potential is reset to the resting potential at $V_0 = 0\,\mathrm{mV}$. 
It remains at this level for short absolute refractory period, $t_\mathrm{ref} = 2\,\mathrm{ms}$, during which all synaptic currents are shunted. 

The response statistics of a LIF neuron, which is driven by randomly arriving input spikes, can be analytically solved in the stationary case. 
Assuming a fixed voltage threshold, $V_\mathrm{th}$, the solution of the first-passage time problem in response to randomly and rapidly fluctuating input yields explicit expressions for the moments of the inter-spike interval distribution \cite{Siegert1951, Ricciardi1977}. 
In particular, the mean response rate of the neuron, $r$, in terms of the mean, $\mu$, and variance, $\sigma^2$, of the fluctuating input is obtained 
\[ 
 r = F(\mu, \sigma) = \left[t_\mathrm{ref} + \tau \sqrt{\pi} \int_{\tilde{V}_0}^{\tilde{V}_\mathrm{th}} \underbrace{e^{u^2} (1+\erf(u))}_{h(u)} \,du \right]^{-1}  
 \label{Eq_mT} 
 \] 
with $\tilde{V}_\mathrm{th} = (V_\mathrm{th}-\mu) / \sigma$ and $\tilde{V}_0 = (V_0 - \mu) / \sigma$. 

Employing a mean field ansatz, the above theory can be applied to networks of identical pulse-coupled LIF neurons, randomly connected with homogeneous in-degrees, and driven by external excitatory input of the same strength. 
Under these conditions, all neurons exhibit the same mean firing rate, which can be determined by a straight-forward self-consistency argument \cite{Amit1997, Brunel2000}: 
The firing rate $r$ is a function of the first two cumulants of the input fluctuations, $\mu$ and $\sigma^2$, which are, in turn, functions of the input. If $s$ is the input (stimulus) firing rate, and $r$ is the mean response rate of all neurons in the network, respectively, we have the relation 
\begin{align} 
  \mu(s,r) &= \tau [J_s s + J_r r N \epsilon (f - g(1-f))], \nonumber 
  \\ \sigma^2(s,r) &= \tau [J_s^2 s + J_r^2 r N \epsilon (f + g^2(1-f))]. 
  \label{Eq_inputMS} 
\end{align} 
Here $J_s$ denotes the amplitude of an excitatory post-synaptic potential (EPSP) of external inputs, and $J_r$ denotes the amplitude of recurrent EPSPs. 
The factor $g$ is the inhibition-excitation ratio, which fixes the strength of inhibitory post-synaptic potentials (IPSPs) to $-gJ_r$. 
Synapses are modeled as $\delta$-functions, where the pre-synaptic current is delivered to the post-synaptic neuron instantaneously, after a fixed transmission delay of $d = 1.5\,\mathrm{ms}$. 

The remaining structural parameters are the total number of neurons in the network, $N$, the connection probability, $\epsilon$, and the fraction $f$ of neurons in the network that are excitatory ($N_\mathrm{exc} = f N$), implying that a fraction $1-f$ is inhibitory ($N_\mathrm{inh} = (1-f) N$). 
For all networks considered here we have used $f = 0.8$ and $g = 8$. 
$J_s$ was always fixed at $0.1\,\mathrm{mV}$. 
For all network connectivities, we fix the in-degree, separately for the excitatory and the inhibitory population, respectively. That is, each neuron, be it excitatory or inhibitory, receives exactly $\epsilon_\mathrm{exc} N_\mathrm{exc}$ connections randomly sampled from the excitatory population and $\epsilon_\mathrm{inh} N_\mathrm{inh}$ connections randomly sampled from the inhibitory population.
Multiple synaptic contacts and self-contacts are excluded. 

In our simulations, inputs are stationary and independent Poisson processes, denoted by a vector $\vect{s}$ of average firing rates. Its $i$-th entry, $s_i$, corresponding to the average firing rate of the input to the $i$-th neuron,  depends on the stimulus orientation $\theta$ and the input preferred orientation (PO) of the neuron $\theta_i^*$ according to 
\[s_i(\theta) = s_b \bigl[ 1 + m \cos(2(\theta - \theta_i^*)) \bigr]. 
\label{Eq_S} 
\] 
The baseline $s_b$ is the level of input common to all orientations, and the peak input is $(1+m) s_b$. 
The input PO is assigned randomly and independently to each neuron in the population. 
To measure the output tuning curves in numerical simulations, we stimulated the networks for $8$ different stimulus orientations, covering the full range between $0\deg$ and $180\deg$ in steps of $22.5\deg$.  
The stimulation at each orientation was run for $15\,\mathrm{s}$, using a simulation time step of $0.1\,\mathrm{ms}$.  
Onset transients (the first $150\,\mathrm{ms}$) were discarded.  


\subsection*{Linearized Rate Equations} 

To quantify the response of a network to tuned input, we first compute its baseline (untuned) output firing rate, $r_b$. 
This procedure is described elsewhere in detail \cite{Sadeh2014}, and we only recapitulate the main steps and equations here. 
If the attenuation of the baseline and amplification of the modulation is performed by two essentially independent processing channels in the network \cite{Sadeh2014}, the baseline firing rate can be computed from the fixed point equation 
\[ 
r_b = F(\mu(s_b,r_b), \sigma(s_b,r_b)), 
\label{Eq_MFrate} 
\] 
the root of which can be found numerically \cite{Brunel2000,Sadeh2014}. 
 
Now we linearize the network dynamics about an operating point defined by the baseline. 
First, we write the full nonlinear rate equation of the network as $\vect{r} = F(\vect{\mu}, \vect{\sigma})$.
Here, the mean and the variance of the input are expressed, in matrix-vector notation, as
\begin{align}
\vect{\mu}(\vect{s}, \vect{r}) = \tau [J_s \vect{s} + W \vect{r}], \nonumber
\\ \vect{\sigma}^2(\vect{s}, \vect{r}) = \tau [J^2_s \vect{s} + V \vect{r}],
\label{Eq_InputVector} 
\end{align}
where $\vect{s}$ and $\vect{r}$ are $N$-dimensional column vectors of input and output firing rates, respectively, and $W$ is the weight matrix of the network.
Its entry $W_{ij}$, the weight of a synaptic connection from neuron $j$ to neuron $i$, is either $0$ if there is no synapse, $J_r$ if there is an excitatory synapse, or $-gJ_r$ if there is an inhibitory synapse. 
Matrix $V$ is the element-wise square of $W$, that is $V_{ij} = W^2_{ij}$.

The extra firing rate of all neurons, $\delta \vect{r} = \vect{r}_m$ (output modulation), in response to a small perturbation of their inputs, $\delta \vect{s} = \vect{s}_m$ (input modulation), is obtained by linearizing the dynamics about the baseline, i.e.\ about $\mu_b$ and $\sigma_b$ (obtained from Eq.~\eqref{Eq_inputMS} evaluated at $r_b$ and $s_b$)
\[
\delta \vect{r} = 
\frac{\partial{F(\mu, \sigma)}}{\partial{\mu}}\bigg|_{\mu_b, \sigma_b} \delta \vect{\mu} + 
\frac{\partial{F(\mu, \sigma)}}{\partial{\sigma}}\bigg|_{\mu_b, \sigma_b} \delta \vect{\sigma}. 
\label{Eq_LinearizedDynamics} 
\]
The partial derivatives of $F(\mu, \sigma)$ at this operating point can be computed from Eq.~\eqref{Eq_mT} as 
\begin{align} 
\alpha = \frac{\partial{F(\mu, \sigma)}}{\partial{\mu}}\bigg|_{\mu_b, \sigma_b} &= 
-F^2(\mu_b, \sigma_b) \tau \sqrt{\pi} 
\frac{\partial}{\partial{\mu}}{\left[\int_{\tilde{V}_0}^{\tilde{V}_\mathrm{th}} h(u)\,du \right]}\bigg|_{\mu_b} \nonumber \\ 
&= -\tau \sqrt{\pi} F^2(\mu_b, \sigma_b) 
\left[ h(\tilde{V}_\mathrm{th}^b) \frac{\partial{\tilde{V}_\mathrm{th}}}{\partial{\mu}}\bigg|_{\mu_b} - 
 h(\tilde{V}_0^b) \frac{\partial{\tilde{V}_0}}{\partial{\mu}}\bigg|_{\mu_b} \right] \nonumber \\ 
&= \frac{\tau \sqrt{\pi}}{\sigma_b} r_b^2 \left[ h(\tilde{V}_\mathrm{th}^b) - h(\tilde{V}_0^b) \right] 
\label{Eq_alpha}
\end{align} 
and, in a similar fashion,
\begin{align} 
\beta = \frac{\partial{F(\mu, \sigma)}}{\partial{\sigma}}\bigg|_{\mu_b, \sigma_b} &= 
-\tau \sqrt{\pi} F^2(\mu_b, \sigma_b) 
\left[ h(\tilde{V}_\mathrm{th}^b) \frac{\partial{\tilde{V}_\mathrm{th}}}{\partial{\sigma}}\bigg|_{\sigma_b} - 
 h(\tilde{V}_0^b) \frac{\partial{\tilde{V}_0}}{\partial{\sigma}}\bigg|_{\sigma_b} \right] \nonumber \\ 
&= \frac{\tau \sqrt{\pi}}{\sigma_b^2} r_b^2 
\left[ h(\tilde{V}_\mathrm{th}^b) (\tilde{V}_\mathrm{th}^b - \mu_b) - 
 h(\tilde{V}_0^b) (\tilde{V}_0^b - \mu_b) \right] 
\label{Eq_beta}
\end{align} 
where $F(\mu_b, \sigma_b) = r_b$, and $\tilde{V}_\mathrm{th}^b$, $\tilde{V}^b_0$, $\mu_b$ and $\sigma_b$ are the corresponding parameters evaluated at the baseline (for further details on this derivation, see \cite{Helias2010}). 

We also need to express $\delta \vect{\mu}$ and $\delta \vect{\sigma}$ in terms of the input perturbations. 
In fact, they can be written in terms of $\delta \vect{s}$ and $\delta \vect{r}$ from Eq.~\eqref{Eq_InputVector} as: 
\begin{align}
\delta \vect{\mu} = \tau [J_s \delta \vect{s} + W \delta \vect{r}], \nonumber \\
\delta \vect{\sigma} = \tau [\frac{J^2_s}{2 \sigma_b} \delta \vect{s} + \frac{V}{2 \sigma_b} \delta \vect{r}]. 
\end{align}

For the total output perturbation, $\vect{r}_m = \delta \vect{r}$, we therefore obtain
\[
\vect{r}_m = \alpha \tau [J_s \vect{s}_m + W \vect{r}_m] + 
\beta \tau [\frac{J^2_s}{2 \sigma_b} \vect{s}_m + \frac{V}{2 \sigma_b} \vect{r}_m].
\label{Eq_vect_rm}
\]
With the simulation parameters used here, our network typically operates in a fluctuation-driven regime of activity with a comparable level of input mean and fluctuations, $\O(\sigma) \approx \O(\mu)$.
As a result, the contribution of the mean, $\alpha \vect{\mu}$, to output modulation in Eq.~\eqref{Eq_vect_rm} is $\O(\sigma \alpha/\beta)$ larger than the contribution of the variance, $\beta \vect{\sigma}$.
In the noise-dominated regime, $\tilde{V}_0^b$ and $\tilde{V}_\mathrm{th}^b$ are small compared to $\mu_b$ in Eq.~\eqref{Eq_beta}, and hence we can write $\beta \approx \frac{-\tau \sqrt{\pi}}{\sigma_b^2} r_b^2 \mu_b \left[ h(\tilde{V}_\mathrm{th}^b)  -  h(\tilde{V}_0^b) \right] $, yielding $\beta \approx - \frac{\mu_b}{\sigma_b} \alpha$.
Thus, with a comparable level of mean and fluctuations, the contribution of the mean to output modulation is $\O(\sigma)$ larger than the contribution of the variance.
In fact, the more the network operates in the noise-dominated regime, the more $\alpha \vect{\mu}$ becomes dominant over $\beta \vect{\sigma}$, making the second term on the right hand side of Eq.~\eqref{Eq_vect_rm} negligible.

For the network shown in Fig.~1, for instance, $r_b \approx 5\,\mathrm{spikes/s}$.
Given the general parameters of our simulation, we obtain $\mu_b = 7\,\mathrm{mV}$ and $\sigma_b = 10\,\mathrm{mV}$.
This yields $\tilde{V}_0^b=-0.7$ and $\tilde{V}_\mathrm{th}^b=1.3$, and finally $\alpha = 1.12$ and $\beta = 0.63$.
In response to feedforward input perturbations, therefore, the contribution of the mean term ($\alpha \tau J_s \vect{s}_m $) is $\frac{2 \alpha \sigma_b}{\beta J_s} \approx 250$ times the contribution of the variance term ($\beta \tau \frac{J^2_s}{2 \sigma_b} \vect{s}_m$).  
In response to recurrent perturbation vectors with zero mean, both the mean term ($\alpha \tau W \vect{r}_m$) and the variance term ($\beta \tau \frac{V}{2 \sigma_b} \vect{r}_m$) would respond with zero output, on average. 
The variance, in contrast, is not zero; a similar computation as in Eq.~\eqref{Eq_inputMS} yields $\tau (\alpha J_r)^2 r N \epsilon [f + g^2(1-f)]$ and $\tau  (\frac{\beta J_r^2}{2 \sigma_b})^2 r N \epsilon [f + g^4(1-f)]$, the terms resulting from the mean and variance contributions, respectively.
That is, the mean contribution is dominant again by a factor of $\frac{4 \sigma_b^2 \alpha [f + g^2(1-f)]}{\beta^2 J_r^2  [f + g^4(1-f)]} \approx 300$.

In the rest of our computation we therefore ignore the second part of the right hand side in Eq.~\eqref{Eq_vect_rm} and approximate the output modulation as:
\[
\vect{r}_m \approx \alpha \tau [J_s \vect{s}_m + W \vect{r}_m].
\label{Eq_vect_rm_simpl}
\]
We call 
\[
\zeta = \tau \alpha =  \frac{\tau^2 \sqrt{\pi}}{\sigma_b} r_b^2 \left[ h(\tilde{V}_\mathrm{th}^b) - h(\tilde{V}_0^b)\right] 
\label{Eq_lin_gain}
\]
 the ``linearized gain'' and write the linearized rate equation of the network in response to small input perturbations as:
\[ 
\vect{r}_m = \zeta W \vect{r}_m + \zeta J_s \vect{s}_m.
\label{Eq_LinRate} 
\] 


\subsection*{Linear and Supralinear Gains}

The gain $\zeta$ is the linearized gain in the firing rate of a single LIF neuron in response to small changes in its mean input, while it is embedded in a recurrent network operating in its baseline AI state. 
That is, the extra firing rate, $\delta r$, of a neuron in response to a perturbation in its input, $\delta s$, when all other neurons are receiving the same, untuned input as before, divided by the input modulation weighted by its effect on the postsynaptic membrane $\zeta = \frac{\delta r}{J_s \delta s}$. 

Alternative to the analytic derivation we pursued above, this gain can also be evaluated numerically by perturbing the baseline firing rate with an extra input, $\delta s$:
\[
\zeta = \frac{\delta r}{J_s \delta s} = 
\frac{\overbrace{F(\mu(s_b + \delta s, r_b), \sigma(s_b + \delta_s , r_b))}^{r_b + \delta r} - \overbrace{F(\mu(s_b,r_b), \sigma(s_b,r_b))}^{r_b}}{J_s \delta s}.
\label{Eq_NumericalGain}
\]
(Note that, as this is the response gain of an individual neuron to an individual perturbation in its input when all other neurons receive the same baseline input, it is not needed to consider the perturbation in the recurrent firing rate, $r$, in the baseline state.)

If this procedure is repeated for each $\delta s$, a numerical $f$--$I$ curve is obtained.
This is the curve we have plotted in Fig.~3A as ``Numerical perturbation''.
If this curve was completely linear, it should not be much different from the results of our analytical perturbation (Eq.~\eqref{Eq_lin_gain}, denoted by ``Linearized gain'' in Fig.~3A).
The results of the numerical perturbation, however, show some supralinear behavior, i.e.\ larger perturbations lead to a higher input-output gain.
As a result, if we compute the gain at a perturbation size equal to the input modulation ($s_m$), a different gain is obtained.
We use the term ``stimulus gain'' to refer to this supralinear gain at the modulation size of input (i.e.\ when $\delta s = s_m$) :
\[
\zeta_s = \frac{\delta r}{J_s s_m} = 
\frac{F(\mu(s_b + s_m, r_b), \sigma(s_b + s_m , r_b)) - F(\mu(s_b,r_b), \sigma(s_b,r_b))}{J_s s_m}.
\label{Eq_StimGain}
\]
This is shown by the red line in Fig.~3A.


\subsection*{Linear Tuning in Recurrent Networks}

Once we obtained the linearized gains at the baseline state of network operation, the linearized rate equation of the network for modulations about the baseline activity is obtained.
Each neuron responds to the aggregate perturbation in its input with a gain obtained by the linearization formalism employed.
The total perturbation consists of a feedforward component, which is the modulation in the input (stimulus) firing rate of the neuron, and a recurrent component, which is a linear sum of the respective output perturbations of the pre-synaptic neurons in the recurrent network.
This can, therefore, be written, in vector-matrix notation, as:
\[ 
\vect{r}_m = \zeta W \vect{r}_m + \zeta J_s \vect{s}_m.
\label{Eq_LinRate} 
\]

If $\one - \zeta W$ is invertible, the output firing rates can be computed directly as 
\[ 
\vect{r}_m = (\one - \zeta \mathbf{W})^{-1} \zeta J_s \vect{s}_m = A \zeta J_s \vect{s}_m, 
\label{Eq_LinRateDir} 
\] 
which can be further expanded into 
\[ 
\vect{r}_m = \sum_{k=0}^{\infty} (\zeta W)^k \zeta J_s \vect{s}_m. 
\label{Eq_LinRateExpanded} 
\] 
Ignoring higher-order contributions $\O((\zeta W)^2)$, Eq.~\eqref{Eq_LinRateExpanded} can be approximated as 
\[ 
\vect{r}_m \approx (\one + \zeta W) \zeta J_s \vect{s}_m. 
\label{Eq_LinearRateSimp} 
\] 
Eq.~\eqref{Eq_LinearRateSimp} for each stimulus orientation returns the modulation of the output firing rate of all neurons in the network in response to a given input modulation. 

We then assume that all inputs $s_i$ are linearly tuned to the stimulus $\vect{\phi}$ according to 
\[s_i(\vect{\phi}) = \psi_i^\ast + \langle \vect{\phi}_i^\ast , \vect{\phi}\rangle, 
\] 
where $\psi_i^\ast$ is the baseline rate in absence of stimulation and the vector $\vect{\phi}_i^\ast$ is the vector of preferred feature for the $i$-th neuron. 
The length of the vector that represents the preferred feature $\|\vect{\phi}_i^\ast\|$ is the tuning strength. 
To ensure the linearity of operation, the firing rate $s_i(\vect{\phi})$ should remain always positive 
\[0 \leq \min_{\vect{\phi}} s_i(\vect{\phi}) = \psi_i^\ast - \|\vect{\phi}_i^\ast\| \, \|\vect{\phi}\| . 
\] 
If this condition is satisfied, the linearity of the tuning and positivity of firing rates remain compatible. If the condition is violated, partial rectification of the neuronal tuning curve follows and the linear analysis does not fully hold. 

To obtain the operation of the network on input preferred feature vectors, we can write Eq.~\eqref{Eq_LinearRateSimp} for input tuning curves 
\begin{align} 
\vect{r}_m(\vect{\phi}) = A \zeta J_s \vect{s}_m(\vect{\phi}) 
                            = A \zeta J_s \vect{\Phi}^\ast \vect{\phi}. 
\label{Eq_LinRate_TC} 
\end{align} 
Here $\vect{\Phi}^\ast$ is a matrix the rows of which are given by the transposed preferred features $(\vect{\phi}_i^\ast)^T$. 
Therefore, all neurons in the recurrent network are again linearly tuned, with preferred features encoded by the rows of the matrix $A J_s \vect{\Phi}^\ast$. 
From here we can compute the matrix of output feature vectors, $\vect{\Phi}_\mathrm{out}^\ast$, as 
\[ 
\vect{\Phi}_\mathrm{out}^\ast = A \zeta J_s \vect{\Phi}^\ast \approx (\one + \zeta W) \zeta J_s \vect{\Phi}^\ast. 
\label{Eq_OutFeat} 
\] 
The first term on the right-hand side is the weighted tuning vector of the feedforward input each neuron receives, 
and the second term is the mixture of tuning vectors of corresponding pre-synaptic neurons in the recurrent network. 


\subsection*{Distribution of Orientation Selectivity} 

The length of the output feature vector represents the amplitude of the modulation component of output tuning curves. 
This is a measure of orientation selectivity, and we compute its distribution here. 

Orientation is a two-dimensional feature, and the input feature vector ($\vect{\Phi}^\ast$ in Eq.\eqref{Eq_OutFeat}) is now a vector of two-dimensional input feature vectors (a vector of vectors). 
Its each entry, corresponding to the input orientation selectivity vector of each neuron, can, therefore, be determined by a length and a direction. 
The length of all vectors is $s_m = m s_b$, as all inputs have the same modulation, 
and the direction is twice the input PO of neurons (see Eq.~\eqref{Eq_S}), which are drawn independently from a uniform distribution on $[0, \pi)$. 
They are assumed to be independent of the weight matrix $W$, implying the absence of feature specific connectivity. 

The feedforward tuning vector of each neuron is accompanied by a contribution from the recurrent network (Eq.~\ref{Eq_OutFeat}). 
For each neuron, the recurrent contribution is a vectorial sum of the input tuning vectors of its pre-synaptic neurons. 
According to the multivariate Central Limit Theorem, the summation of a large number of independent random variables leads to an approximate multi-variate normal distribution of the output features. 
Tuning strength is given by the length of output tuning vectors, $L = \|\phi\| = \sqrt{{\phi_{\mathrm{out}, x}^\ast}^2 +{\phi_{\mathrm{out}, y}^\ast}^2}$. 
For a bivariate normal distribution with parameters $\mu_L$ and $\sigma^2_L$, we can compute the distribution of this length 
\[ 
\Lambda_2(L;\mu_L,\sigma_L^2) = \frac{L}{\sigma_L^2} e^{-(L^2 + \mu_L^2)/(2\sigma_L^2)} I_0 (\frac{L \mu_L}{\sigma_L^2}), 
  \label{Tot_tuning} 
\] 
where 
\[ 
I_0(z) = \frac{1}{\pi} \int_0^\pi e^{z \cos(\psi)} d\psi \nonumber 
\] 
is the modified Bessel function of the first kind and zeroth order. 
Therefore, we only need to compute the mean and the variance of the resulting distribution. 

The mean of the distribution $\mu_L$ is equal to the length of feedforward feature vector, $\zeta J_s s_m$. 
This is because the expected value of the contribution of the recurrent network vanishes in each direction 
\begin{align} 
\E[\Phi_{\mathrm{rec}, x}^\ast] 
= \E[\zeta W \zeta J_s s_m \cos(2 \Theta^\ast)] 
= \zeta^2 J_s s_m \E[W] \E[\cos(2\Theta^\ast)] = 0. 
\end{align} 
$W$ and $\Theta^\ast$ denote, respectively, the random variables from which the weights and input POs are drawn. 
A similar computation yields $\E[\Phi_{\mathrm{rec}, y}^\ast] = 0$. 
Here we have used the property that the two random variables $W$ and $\Theta^\ast$ are independent, 
and that all orientations are uniformly represented in the input ($\E[\cos(2\Theta^\ast)] = 0$). 
As a result, we obtain 
\[ 
\E[L] = \E[\Phi_{\mathrm{ffw}}^\ast + \Phi_{\mathrm{rec}}^\ast] = \zeta J_s s_m. 
\] 

The recurrent contribution does not, on average, change the length of output feature vectors. 
However, it creates a distribution of selectivity, which can be quantified by its variance 
\begin{align} 
\Var[\Phi_{\mathrm{rec}, x}^\ast]  &= (\zeta^2 J_s s_m)^2 \Var[W]  \Var[\cos(2\Theta)] \nonumber \\ 
           &= (\zeta^2 J_s s_m)^2 \Var[W]  \int_0^\pi{\cos^2(2\Theta) d\Theta} \nonumber \\ 
           &= \frac{1}{2} (\zeta^2 J_s s_m)^2 \Var[W]. 
\end{align} 
Again, we have exploited the independence of random variables $W$ and $\Theta$, and the uniform representation of input POs ($\E[\cos(2\Theta^\ast)] = 0$), to factorize the variance, 
i.e.\ $\Var[W \cos(2\Theta)] = \Var[W]  \Var[\cos(2\Theta)]$. 
Similar computation yields the same variance for the second dimension. 

For our random networks, the weights for each row of the weight matrix are drawn from a binomial distribution, $W$. 
The number of non-zero elements is determined by connection probabilities ($\epsilon_\mathrm{exc}$ and $\epsilon_\mathrm{inh}$ for excitation and inhibition respectively), and each non-zero entry is weighted by the synaptic strength ($J_r$ and $-gJ_r$ for excitation and inhibition respectively). 
The variance $\Var[W]$ can therefore be computed explicitly: 
\[ 
\Var[W] = J_r^2 N \epsilon (1-\epsilon) \left[ f + g^2(1-f) \right]. 
\] 
For more complex connectivities, the variance can be numerically computed from the weight matrix. 
For our networks here, the mean and the variance of the distribution of output tuning vectors can, therefore, be expressed as 
\[ 
\mu_L = \zeta J_s s_M \quad \mathrm{and} \quad 
\sigma^2_L = \frac{1}{2} (\zeta^2 J_s s_M)^2 J_r^2 N \epsilon (1-\epsilon) \left[ f + g^2(1-f) \right]. 
\label{Eq_MeanSigDist} 
\] 

For an output tuning curve with a cosine shape, $R(\theta) = R_b + R_m \cos(\theta - \theta^*)$, the tuning strength we introduced above corresponds to $R_m$, namely the modulation (F2) component of the tuning curve. 
$R_b$ is also obtained as the baseline firing rate of the network, $r_b$, from Eq.~\eqref{Eq_MFrate}. 
To compare the prediction with the result of our simulations, we compute the mean and modulation of individual output tuning curves from the simulated data. 
Mean and modulation are taken from the zeroth and the second Fourier components of each tuning curve (F0 and F2 components), respectively. 
The distribution given by Eq.~\eqref{Tot_tuning} should, therefore, precisely match the distribution of modulation (F2) component of output tuning curves obtained from simulations, if our linear analysis grasps the essential mechanisms of orientation selectivity in model recurrent networks.

\section*{Results} 

\subsection*{Erd\H{o}s-R\'{e}nyi Random Networks} 

\begin{figure} 

\centering\includegraphics[width=6.0in]{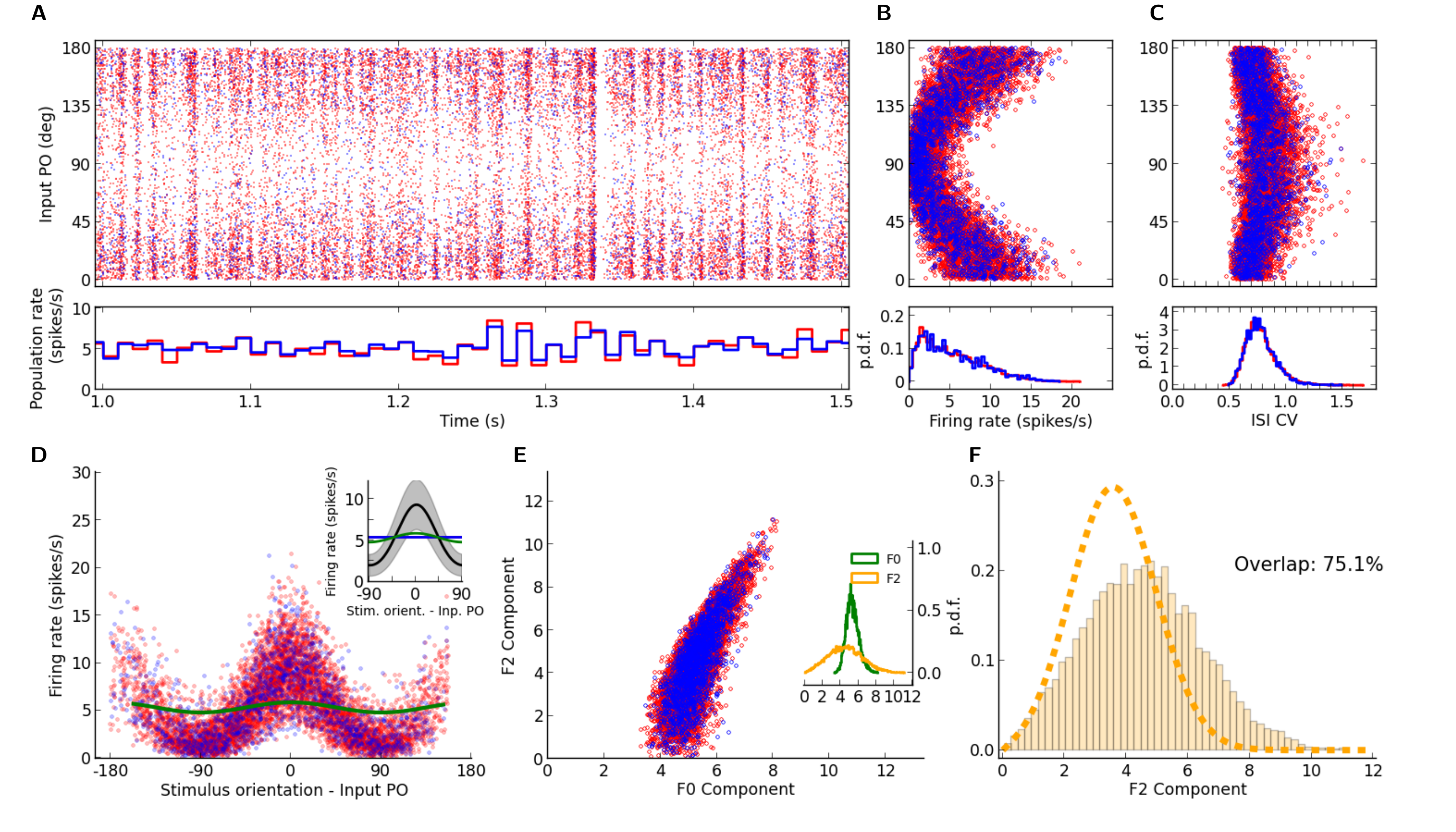} 

\caption{{\bf Distribution of orientation selectivity in networks with Erd\H{o}s-R\'{e}nyi random connectivity.} 
\textbf{(A)}~Raster plot of network activity in response to stimulus with orientation $\theta = 0\deg$. 
Neurons are sorted according to their input preferred orientations, $\theta^\ast$, indicated on the vertical axis. 
The histogram on the bottom shows the population firing rates, averaged in time bins of $10\,\mathrm{ms}$ width. 
Here, and in all other figures, red and blue colors denote excitatory and inhibitory neurons, or neuronal populations, respectively. 
\textbf{(B)}~Average firing rates, for all neurons in the network, estimated from the spike count over the whole stimulation period ($t_\mathrm{stim} = 15\,\mathrm{s}$). 
The distribution of firing rates over the population is depicted in the histogram at the bottom. 
\textbf{(C)}~Coefficient of Variation (CV) of the inter-spike intervals (ISI), $\mathrm{CV}_\mathrm{ISI} = \mathrm{std}(\mathrm{ISI})/\mathrm{mean}(\mathrm{ISI})$, computed for all neurons in the network with more than $10$ spikes during the stimulation. 
The distribution of $\mathrm{CV}_\mathrm{ISI}$ is plotted at the bottom. 
\textbf{(D)}~Sample output tuning of $800$ excitatory and $200$ inhibitory neurons randomly chosen from the network, all aligned at their input preferred orientations. 
The input tuning (green, same as Eq.~\eqref{Eq_S}) is normalized to the population average of the baseline (mean over all orientations) of output tuning curves. 
Inset: The mean (across population) of aligned output tunings are shown in black.
The gray shading indicates $\mathrm{mean} \pm \mathrm{std}$ extracted from the population. 
Linearly interpolated versions of individual tuning curves (generated at a resolution of $1\deg$) have been used to compute $\mathrm{mean}$ and $\mathrm{std}$ of aligned tuning curves. 
The population average of the baseline (mean over all orientations) of output tuning curves is shown separately for excitatory and inhibitory populations with a red and a blue line, respectively (the lines highly overlap, since the average activity almost coincide for both populations).
The normalized input tuning curve (green) is obtained by the same method as used for the main plot.
\textbf{(E)}~Scatter plot of F0 and F2 components, extracted from individual output tuning curves in the network. 
The individual distributions of F0 and F2 components over the population are plotted in the inset. 
\textbf{(F)}~Distribution of single-neuron F2 components from a network simulation (histogram) compared with the prediction of our theory (dashed line, computed from Eq.~\eqref{Tot_tuning}). 
To evaluate the goodness of match, the overlap of the empirical and predicted probability density functions ($\mathrm{Pr}_\mathrm{emp}$ and $\mathrm{Pr}_\mathrm{prd}$, respectively) is computed as $\int_{-\infty}^{\infty} \mathrm{min}(\mathrm{Pr}_\mathrm{emp}(x'),\mathrm{Pr}_\mathrm{prd}(x'))\,dx'$. 
This returns an overlap index between $0\%$ and $100\%$, corresponding to no overlap and perfect match of distributions, respectively. 
Parameters of the network simulation are: $N = 10\,000$, $\epsilon_\mathrm{exc} = \epsilon_\mathrm{inh} = 10\%$, $J_r = 0.25\,\mathrm{mV}$, $g = 8$, $s_b = 15\,000\,\mathrm{spikes}/\mathrm{s}$, $J_s = 0.1\,\mathrm{mV}$, $m = 10\%$. 
} 
\label{Fig_ER} 
\end{figure}

We first study excitatory-inhibitory Erd\H{o}s-R\'{e}nyi random networks of LIF neurons (Eq.~\eqref{Eq_NeuronModel}) with a doubly fixed in-degree, namely where both the excitatory in-degree and the inhibitory in-degree is fixed for both excitatory and inhibitory neurons.
Figs.~\ref{Fig_ER}A--C show the response of a network with $J_r = 0.25~\mathrm{mV}$ and $\epsilon_\mathrm{exc} = \epsilon_\mathrm{inh} = 0.1$ to the stimulus of $0\,\deg$ orientation. 
The network with these parameters operates in the fluctuation-driven regime, which shows asynchronous-irregular (AI) dynamics (Fig.~\ref{Fig_ER}A), with low firing rates (Fig.~\ref{Fig_ER}B) and high variance of inter-spike intervals (ISI) (Fig.~\ref{Fig_ER}C). 
The network at this regime is capable of amplifying the weak tuning of the input, as it is reflected both in the network tuning curve in response to one orientation  (Fig.~\ref{Fig_ER}B) and in individual tuning curves in response to different stimulus orientations  (Fig.~\ref{Fig_ER}D). 

The joint distribution of the modulation (F2) component of (individual) output tuning curves and the respective baseline (F0) component (Fig.~\ref{Fig_ER}E) shows that the average values of these two components have become comparable after network operation. 
However, the F2 component has a much broader distribution (Fig.~\ref{Fig_ER}E, inset). 
The distribution predicted by our theory (Eq.~\eqref{Tot_tuning}) matches partially with the distribution measured in the simulations (Fig.~\ref{Fig_ER}F). The degree of match is quantified by an index, which assesses the overlap area of the two probability distributions. 



As our analysis is based on the assumption of linearity of network interactions, the result of our theoretical prediction holds only if the network is operating in the linear regime. 
Any violation of our linear scheme would, therefore, lead to a deviation of the linear prediction from the measured distribution. 
The remaining discrepancy should, therefore, be attributed to any factor which
invalidates our approximation scheme here.
Possible contributing factors of this sort in our networks are partial rectification of tuning curves, correlations and synchrony in the network providing the input, and supralinearity of neuronal gains. 

Partial rectification of firing rates is obvious in Fig.~\ref{Fig_ER}B. 
However, this does not seem to be a very prominent effect. Only a small fraction of the population is strictly silent, as is evident in the distribution of firing rates (Fig.~\ref{Fig_ER}B, bottom). 
Correlations, in contrast, seems to be a more important contributor, as is reflected in the raster plot of network activity (Fig.~\ref{Fig_ER}A). 

\begin{figure}  

\centering\includegraphics[width=6.0in]{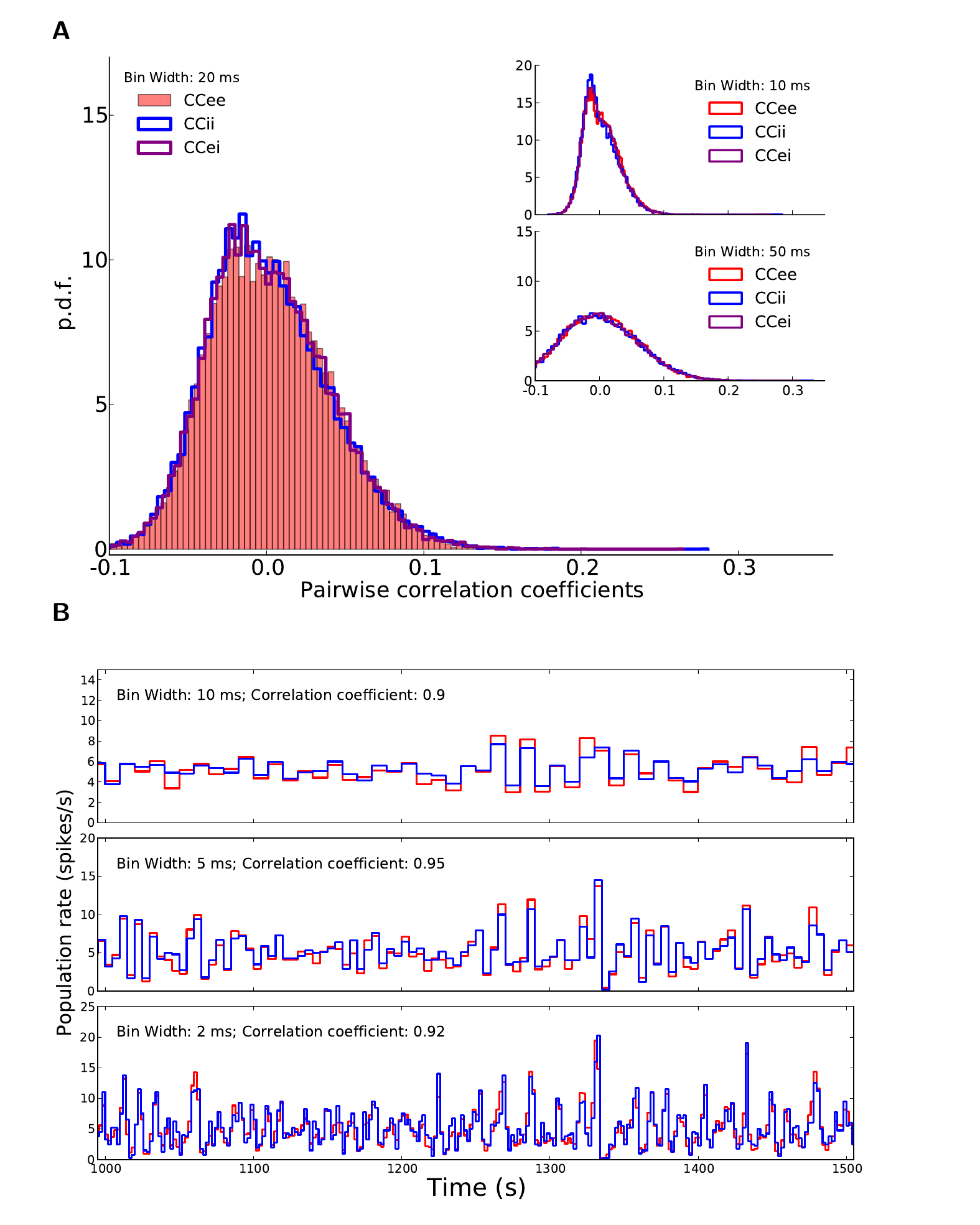} 

\caption{{\bf Correlations in the network.} 
\textbf{(A)}~Distribution of correlation coefficients for pairs of neurons in the network. 
For the example network of Fig.~\ref{Fig_ER}, the distribution of Pearson correlation coefficients (CC) between spike trains of pairs of neurons is plotted. 
$200$ excitatory and $200$ inhibitory neurons are randomly sampled from the network and all pairwise correlations (between pairs of excitatory, $\mathrm{CC}_\mathrm{ee}$, between pairs of inhibitory, $\mathrm{CC}_\mathrm{ii}$, and between excitatory and inhibitory, $\mathrm{CC}_\mathrm{ei}$, samples), based on spike counts in bins of width $20\,\mathrm{ms}$ are computed. 
The corresponding distributions for smaller ($10\,\mathrm{ms}$) and larger ($50\,\mathrm{ms}$) bins are shown in the inset (top and bottom, respectively). 
\textbf{(B)}~The time series for the excitatory and inhibitory population spike counts indicate a fine balance on the population level. 
The correlation of activity between excitatory (red) and inhibitory (blue) populations is quite high on different time scales. 
The similarity of the temporal pattern of population activities is again quantified by the Pearson correlation coefficient. 
\label{Fig_ER_Corr} 
} 
\end{figure}

To investigate the possible contribution of correlations in the distribution of orientation selectivity, we plotted the distribution of pairwise correlations in the network (Fig.~\ref{Fig_ER_Corr}). 
Although the distribution of pairwise correlations has a very long tail, on average correlations are very small in the network (Fig.~\ref{Fig_ER_Corr}A). 
This is the case for excitatory-excitatory, excitatory-inhibitory, and inhibitory-inhibitory correlations, and there is the same trend when spike counts are computed for different bin widths (Fig.~\ref{Fig_ER_Corr}A, insets). 
Low pairwise correlations in the network are a result of recurrent inhibitory feedback, which actively decorrelates the network activity \cite{Renart2010, Pernice2011, Tetzlaff2012}. 
As illustrated in Fig.~\ref{Fig_ER_Corr}B, upsurges in the population activity of excitatory neurons are tightly coupled to a corresponding increase in the activity of the inhibitory population. However, the cancellation is not always exact and some residual correlations remain. 



Since each neuron receives random inputs from $10\%$ of the population, approximately the same correlation of excitation and inhibition is, on average, also expected in the recurrent input to each neuron.Note that, as our networks are inhibition-dominated, the net recurrent inhibition would be stronger than the net recurrent excitation (indeed twice as strong, given the parameters we have used). 
Altogether, this implies that inhibition is capable of fast tracking of excitatory upsurges (Fig.~\ref{Fig_ER_Corr}B) such that fast fluctuations in the population activity would not be seen in the recurrent input from the network. 

\begin{figure}   

\centering\includegraphics[width=6.0in]{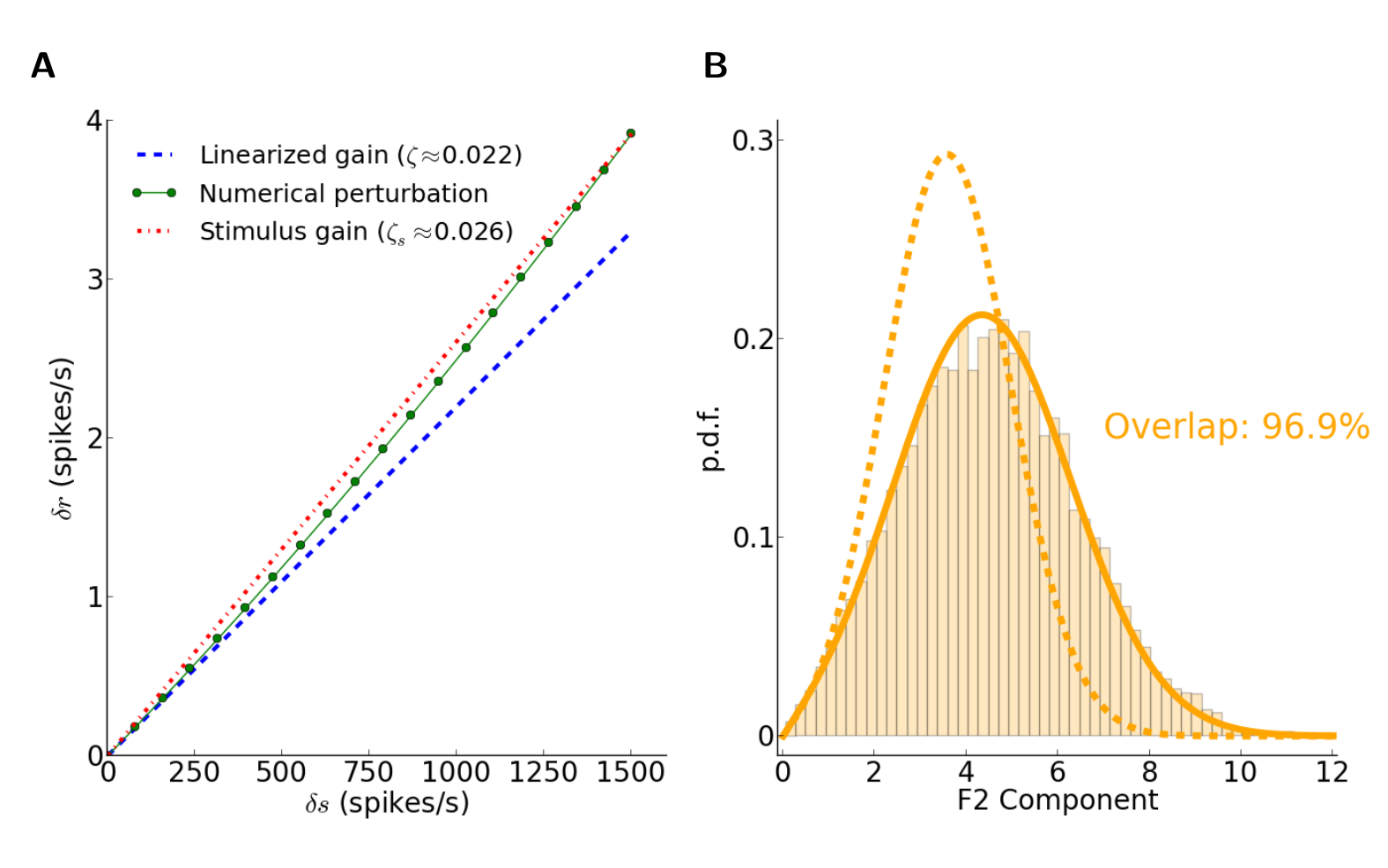} 

\caption{{\bf Supralinear neuronal gain affects the linear prediction.} 
\textbf{(A)}~Discrepancy of the linearized gain with the gain computed at stronger input modulations. 
The linearized gain of the neuron obtained analytically from Eq.~\eqref{Eq_lin_gain} (dashed blue line) is compared with the numerical solution of Eq.~\eqref{Eq_MFrate} with an input perturbation equal to the modulation in the feedforward input, $\delta s = s_m = m s_b = 1\,500\,\mathrm{spikes}/\mathrm{s}$ (see Eq.~\eqref{Eq_NumericalGain} in Methods). 
The red line shows the corresponding linearized gain that would have been computed with this perturbation, $\zeta_s$ (Eq.~\eqref{Eq_StimGain}). 
\textbf{(B)}~Comparison of our theoretical prediction of the distribution with $\zeta$ and $\zeta_s$ (dashed and solid lines, respectively). 
The overlap index of the improved prediction, i.e.\ when $\zeta$ is replaced by $\zeta_s$ in Eq.~\eqref{Eq_MeanSigDist}, has greatly increased. 
}  
\label{Fig_ER_StimGain} 
\end{figure}

Finally, the single-neuron gain that we computed by linearization (Eq.~\eqref{Eq_lin_gain}) could be a source of mismatch, as for a highly non-linear system it might only be valid for small perturbations in the input, and not for stronger modulations. 
This is shown in Fig.~\ref{Fig_ER_StimGain}A, where the linearized gain, $\zeta$, from Eq.~\eqref{Eq_lin_gain} is compared with $\zeta_s$, the numerically obtained neuronal gain (see Eq.~\eqref{Eq_NumericalGain} in Methods) when the perturbation has the size of the input modulation, $s_m = m s_b$. 
This gain could be approximated analytically by expanding Eq.~\eqref{Eq_MFrate} to higher order terms. Here, however, we have computed this gain numerically (Eq.~\eqref{Eq_StimGain}). 



When the prediction of Eq.~\eqref{Tot_tuning} is repeated with the new gain ($\zeta_s$), a great improvement in the match between the measured and predicted distributions is indeed observed (Fig.~\ref{Fig_ER_StimGain}B). 
We therefore concluded that the main source of mismatch in our prediction was our misestimate of the actual neuronal gains. 
Other sources of nonlinearity, like rectification and correlations, could therefore be responsible for the remaining discrepancy of distributions (less than $5\%$ in the regime considered here). 
However, given so many possible sources of nonlinearity in our networks, both at the level of spiking neurons and network interactions, it is indeed quite surprising that a linear prediction works so well. 

A remark about rectification in our networks should be made at this point. 
In the type of networks we are considering here, rectification is in fact not a single-neuron property, i.e.\ only the result of a rectification effect due to the spike threshold in the LIF neuron. 
This is not the case as the linearized gain of neurons within the network (Eq.~\eqref{Eq_lin_gain}) implies a non-zero response even to small perturbations in the input. 
This is a result of (internally generated) noise within the recurrent network, as a consequence of balance of excitation and inhibition, which smoothens the embedded $f$-$I$ curve \cite{Hansel2002, Miller2002}. 
Rectification could therefore only happen at the level of network, e.g.\ by increasing the amount of inhibition. 

\begin{figure}   

\centering\includegraphics[width=6.0in]{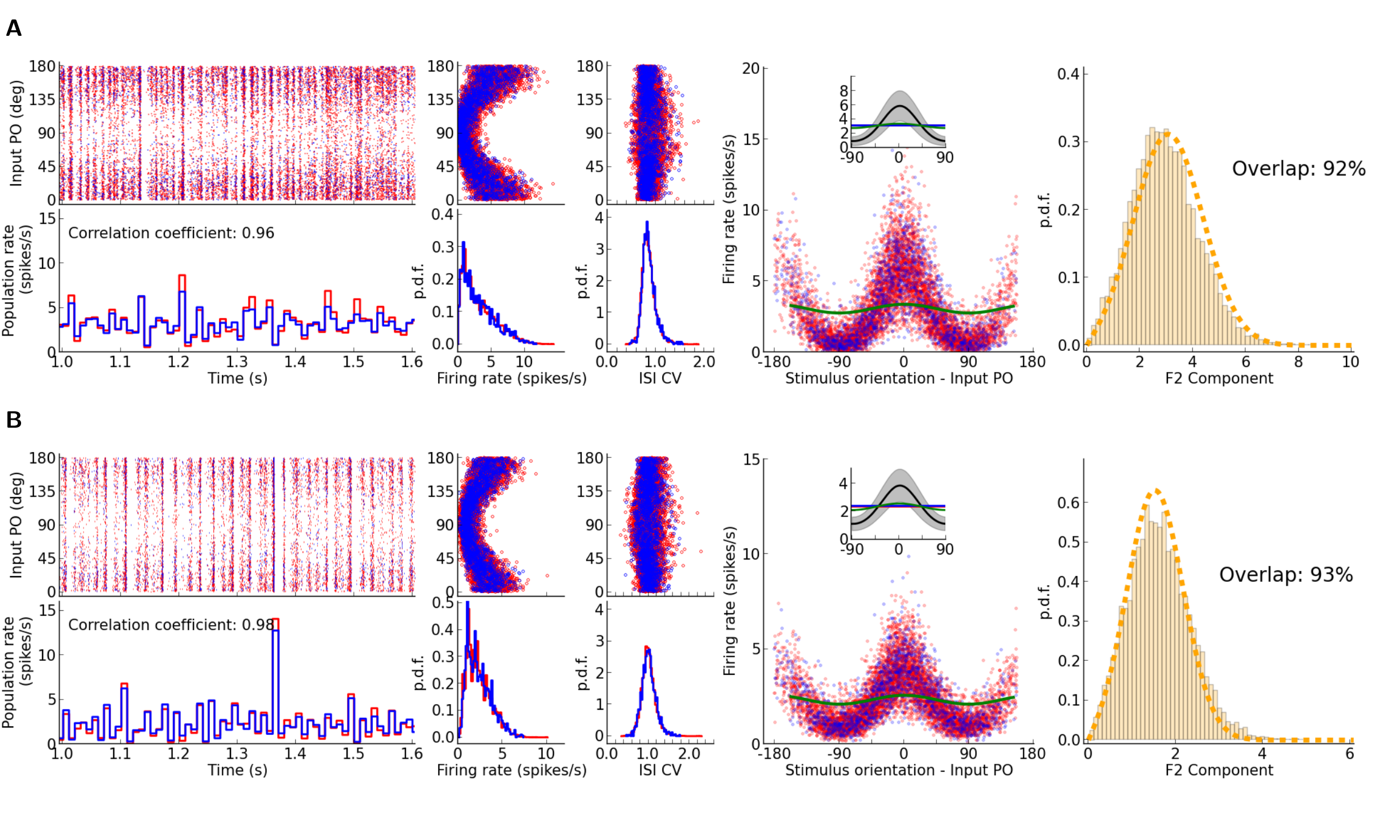} 
 
\caption{{\bf The impact of the strength of recurrent coupling on the distribution of selectivities.} 
The figure layout is similar to Fig.~\ref{Fig_ER} (panel (E) not included), shown are networks with stronger recurrent couplings. 
In (A), the recurrent coupling is increased by doubling the connection density; 
in (B), this is further enhanced by doubling all recurrent weights. 
The parameters of network simulations are: 
\textbf{(A)} $N = 10\,000$, $\epsilon_\mathrm{exc} = \epsilon_\mathrm{inh} = 20\%$, $J_r = 0.25\,\mathrm{mV}$, $g = 8$, $s_b = 15\,000\,\mathrm{spikes}/\mathrm{s}$, $J_s = 0.1\,\mathrm{mV}$, $m = 10\%$, and 
\textbf{(B)} $N = 10\,000$, $\epsilon_\mathrm{exc} = \epsilon_\mathrm{inh} = 20\%$, $J_r = 0.5\,\mathrm{mV}$, $g = 8$, $s_b = 15\,000\,\mathrm{spikes}/\mathrm{s}$, $J_s = 0.1\,\mathrm{mV}$, $m = 10\%$. 
The predicted distributions are computed by considering $\zeta_s$ (see Fig.~\ref{Fig_ER_StimGain}). 
} 
\label{Fig_ER_Rec} 
\end{figure}

As our networks are inhibition-dominated, increasing the recurrent coupling would be one way to increase the inhibitory feedback within the network. 
This can be done in two different ways, either by increasing the connection density or by increasing the weights of synaptic connections. 
The first strategy is tried in Fig.~\ref{Fig_ER_Rec}A, where the connection probability has been increased (from $\epsilon_\mathrm{exc} = \epsilon_\mathrm{inh} = 0.1$ to $\epsilon_\mathrm{exc} = \epsilon_\mathrm{inh} = 0.2$). 
The second strategy is added to the first in Fig.~\ref{Fig_ER_Rec}B, where an increase in the connection density is accompanied by an increase in synaptic weights (from $J_r = 0.25~\mathrm{mV}$ to $J_r = 0.5~\mathrm{mV}$). 
In both cases, however, a significant rectification of tuning curves did not result, and the prediction of our linear theory still holds. 

This unexpected effect can be explained intuitively as follows: 
An increase in recurrent coupling not only decreases the baseline firing rate of the network, but also changes neuronal gains ($\zeta$ and $\zeta_s$). 
A crucial factor in determining this gain is the average membrane potential of neurons in the network, which in turn sets the mean distance to threshold. The larger the mean distance to threshold is in the network, the less is the neuronal gain. 
This in turn decreases the mean F2 component of output tuning curves. As a result, with a reduced baseline firing rate, a significant rectification of tuning curves still does not follow, as output modulation components have been scaled down by a comparable factor. 
This is indeed the case in networks of Fig.~\ref{Fig_ER_Rec}, where the mean (over neurons) membrane potential (temporally averaged) and the neuronal gains have both been decreased compared to the network of Fig.~\ref{Fig_ER} (results not shown; for a detailed analysis, see \cite{Sadeh2014}). 



\subsection*{Networks With Distance-Dependent Connectivity} 

\begin{figure}   

\centering\includegraphics[width=6.0in]{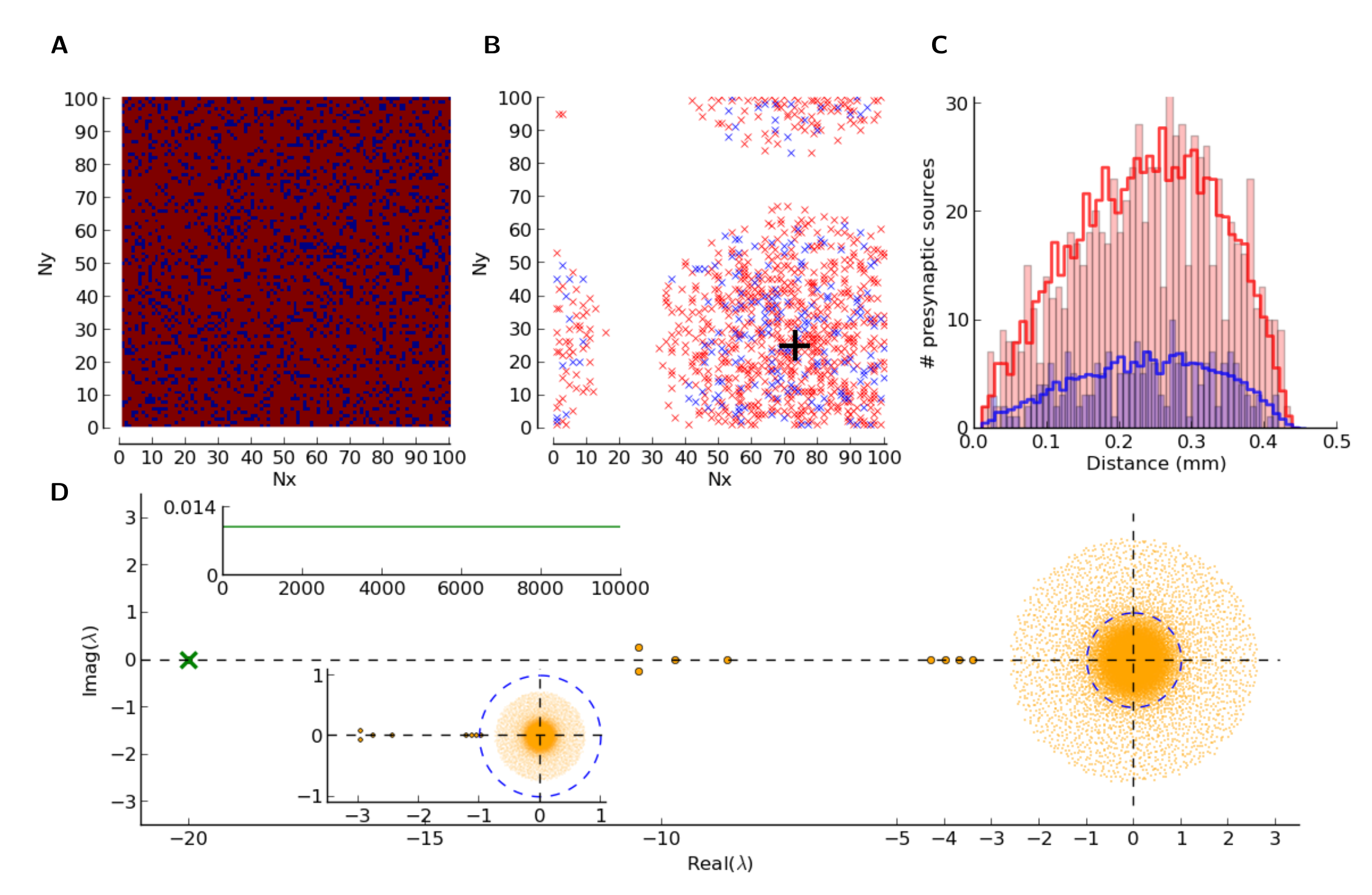} 

\caption{{\bf Networks with distance-dependent connectivity.} 
\textbf{(A)}~Random positioning of $N_\mathrm{exc} = 8\,000$ excitatory (red) and $N_\mathrm{inh} = 2\,000$ inhibitory (blue) neurons in a square, representing a flat $1\,\mathrm{mm} \times 1\,\mathrm{mm}$ sheet of cortex, wrapped to a torus. 
\textbf{(B)}~For a sample (excitatory) neuron (large black cross), positions of excitatory (red) and inhibitory (blue) pre-synaptic neurons are explicitly shown as little crosses. 
A Gaussian connectivity profile with $\sigma_\mathrm{exc} = \sigma_\mathrm{inh} = 0.55\,\mathrm{mm}$ was assumed. 
For each post-synaptic neuron, we fixed the number of randomly drawn pre-synaptic connections of either type, i.e.\ $C_\mathrm{exc} = \epsilon_\mathrm{exc} N_\mathrm{exc}$ and $C_\mathrm{inh} = \epsilon_\mathrm{inh} N_\mathrm{inh}$ ($\epsilon_\mathrm{exc} = \epsilon_\mathrm{inh} = 10\%$). 
Multiple synapses and self-coupling were not allowed. 
\textbf{(C)}~Histogram of distances to pre-synaptic neurons for the sample neuron (bars) and for the entire population (lines). 
\textbf{(D)}~Eigenvalue spectrum of the weight matrix, $W$. 
Weights are normalized by the reset voltage, $V_\mathrm{reset} = V_\mathrm{th} - V_0 = V_\mathrm{th}$, leading to $w_{ij} = J_r/V_\mathrm{th}$ or $-g J_r/V_\mathrm{th}$, depending on whether the synapse is excitatory or inhibitory, respectively. We used $J_r = 0.5\,\mathrm{mV}$. 
For better visibility, the eigenvalues outside the bulk of the spectrum are shown by larger dots. 
The green cross marks the eigenvalue corresponding to the uniform eigenmode, which is plotted in the top inset. 
Re-normalized spectrum, according to the gain $\zeta_s$, is shown in the bottom inset; i.e.\ $w_{ij} = \zeta_s J_r$ and $-g \zeta_s J_r$, for excitatory and inhibitory connections, respectively. 
} 
\label{Fig_DistDepCon} 
\end{figure}

To extend the scope of the linear analysis, we asked if our theory can also account for networks with different statistically defined topologies. 
In particular, we considered networks with a more realistic pattern of distance-dependent connectivity: 
Each neuron is assigned a random position in a two-dimensional rectangle representing a $1\,\mathrm{mm} \times 1\,\mathrm{mm}$ flat sheet of cortex (Fig.~\ref{Fig_DistDepCon}A). 
The probability of having a connection between a pre-synaptic excitatory (inhibitory) neuron to a given post-synaptic neuron falls off as a Gaussian function with distance, with parameter $\sigma_\mathrm{exc}$ ($\sigma_\mathrm{inh}$). 
Similar to the Erd\H{o}s-R\'{e}nyi random networks considered before, we fix the in-degree, i.e.\ each neuron receives exactly $\epsilon_\mathrm{exc} N_\mathrm{exc}$ excitatory and $\epsilon_\mathrm{inh} N_\mathrm{inh}$ inhibitory connections. 
Multiple synaptic contacts and self-contacts are not allowed. 

The connectivity profile is illustrated in Figs.~\ref{Fig_DistDepCon}B, C. 
The pre-synaptic sources of a sample neuron are plotted in Fig.~\ref{Fig_DistDepCon}B, for $\sigma_\mathrm{exc} = \sigma_\mathrm{inh} = 0.55\,\mathrm{mm}$. 
The resulting distribution of the distances of connected neurons, for the example neuron and for the entire population, is shown in Fig.~\ref{Fig_DistDepCon}C. 

Note that the connectivity depends only on the physical distance. As input preferred orientations are assigned randomly and independently of the actual position of neurons in space, distance-dependent connectivity does not imply any feature-specific connectivity. That is, neither a spatial nor a functional map of orientation selectivity is present here.

Before discussing the simulations of the spiking networks, it is informative to look at the eigenvalue spectrum of the associated weight matrix, $W$. 
It is plotted, for $J_r = 0.5\,\mathrm{mV}$ and $\epsilon_\mathrm{exc} = \epsilon_\mathrm{inh} = 10\%$, in Fig.~\ref{Fig_DistDepCon}D. 
Each entry of the matrix is normalized by the reset voltage, $V_\mathrm{reset} = V_\mathrm{th} - V_0 = V_\mathrm{th}$, for the eigenvalue spectrum shown in the main panel. 
The effective firing rate equation of the network can then be written as $\tau d{\vec{r}}/d{t} = -\vec{r} + \frac{1}{V_\mathrm{th}} [W \vec{r} + J_s \vec{s}]$. 
The exceptional eigenvalue (green cross) corresponding to the uniform eigenvector (inset, top) and the bulk of eigenvalues (orange dots) are the structural properties that this network has in common with the previous Erd\H{o}s-R\'{e}nyi networks (not shown). 
There is, however, a small number of additional (in this case, $8$) eigenvalues in between, which are the consequence of the specific realization of our distance-dependent connectivity here. 
The corresponding eigenmodes will, in principle, affect the response of the network, both in its spontaneous state and in response to stimulation. 

All these eigenvalues have, however, negative real parts. 
They will, therefore, ensure the stability of the linearized network dynamics, as far as these eigenmodes are concerned. 
The bulk of the spectrum, in contrast, also comprises eigenvalues with real parts larger than $1$, which implies an instability. 
An alternative normalization of the weight matrix according to the neuronal gain $\zeta_s$ (Fig.~\ref{Fig_DistDepCon}, inset, bottom; see also \cite{Sadeh2014}), however, does not render these modes unstable. 

Here, we are resorting to a linearized rate equation describing the response of the network to (small) perturbations, 
$\tau d{\vec{r}}/d{t} = -\vec{r} + \zeta_s [W \vec{r} + J_s \vec{s}]$ (see Eq.~\eqref{Eq_LinRate} in Methods). 
The eigendynamics corresponding to the common-mode (green cross) is faster, and hence it relaxes to the fixed point more rapidly than the other eigenmodes. 
The common mode effectively leads to the uniform, baseline state of the network (reflected in the baseline firing rate, $r_b$), about which the network dynamics has indeed been linearized in our linear prediction. 
The effect of other eigenmodes, in the stationary state, should therefore be computed by considering the linearized gain about that uniform, baseline state. 



\begin{figure}   

\centering\includegraphics[width=6.0in]{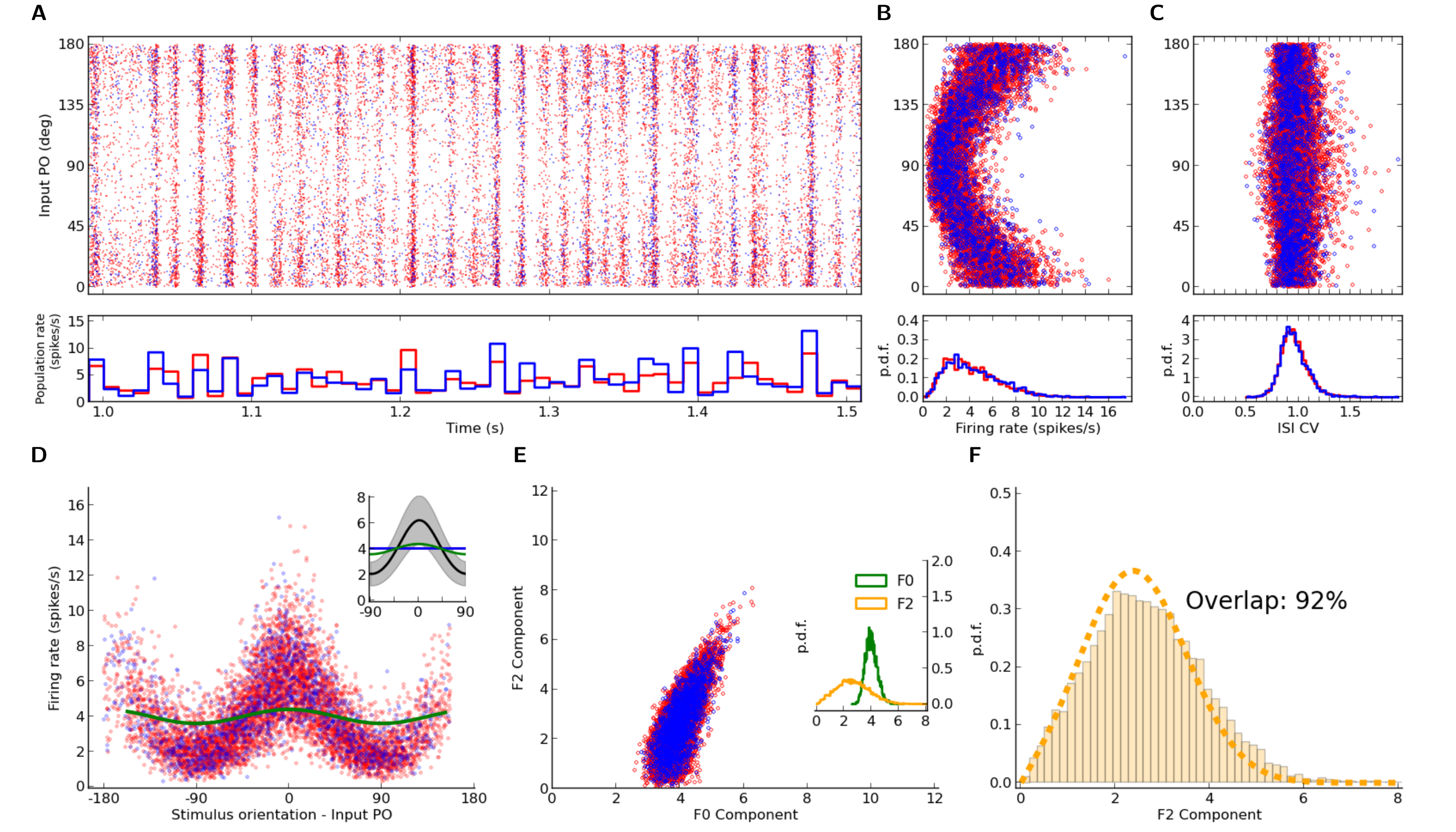} 

\caption{{\bf Distribution of orientation selectivity in a network with distance-dependent connectivity.} 
Same figure layout as Fig.~\ref{Fig_ER}, for a network with distance-dependent connectivity, similar to Fig.~\ref{Fig_DistDepCon}. 
Parameters of the network simulation are: $N = 10\,000$, $\epsilon_\mathrm{exc} = \epsilon_\mathrm{inh} = 10\%$, $\sigma_\mathrm{exc} = \sigma_\mathrm{inh} = 0.55\,\mathrm{mm}$, $J_r = 0.5\,\mathrm{mV}$, $g = 8$, $s_b = 15\,000\,\mathrm{spikes}/\mathrm{s}$, $J_s = 0.1\,\mathrm{mV}$, $m = 10\%$. 
Note that the distribution of F2 components is computed by using the stimulus gain $\zeta_s$, as in Fig.~\ref{Fig_ER_StimGain}. 
} 
\label{Fig_DDC} 
\end{figure}

Simulation results for a network with this connectivity are illustrated in Fig.~\ref{Fig_DDC}. 
Inspection of the spiking activity of the network (Fig.~\ref{Fig_DDC}A) does not suggest a behavior very different from the behavior of random networks shown in Fig.~\ref{Fig_ER}. 
The irregularity of firing is, however, more pronounced, as the variance of inter-spike intervals is larger (Fig.~\ref{Fig_DDC}C); 
the ISI CV has indeed a distribution about $1$, which is more similar to the strongly coupled networks described in Fig.~\ref{Fig_ER_Rec}. 

Similar to Erd\H{o}s-R\'{e}nyi networks, networks with distance-dependent local connectivity are capable of amplifying the weak tuning of the input signal, and comparable levels of baseline (F0) and modulation (F2) components are emerging (Fig.~\ref{Fig_DDC}E). 
When the predicted distribution of F2 components is obtained applying the normalization by the linear gain $\zeta_s$, a very good match to the measured distribution is obtained (Fig.~\ref{Fig_DDC}F), comparable to predictions in Fig.~\ref{Fig_ER_Rec}, and only slightly worse than the prediction in Fig.~\ref{Fig_ER}. 



Although partial rectification of tuning curves seems to be negligible in the example shown (Fig.~\ref{Fig_DDC}B), correlations in the network could still be responsible for the remaining discrepancy. 
Moreover, size and structure of correlations in the network might be different here as compared to random networks due to non-homogeneous connectivity. 
Distance-dependent connectivity implies that connectivity is locally dense, which can lead to more shared input and this way impose strong correlations at the output. 

\begin{figure}  

\centering\includegraphics[width=6.0in]{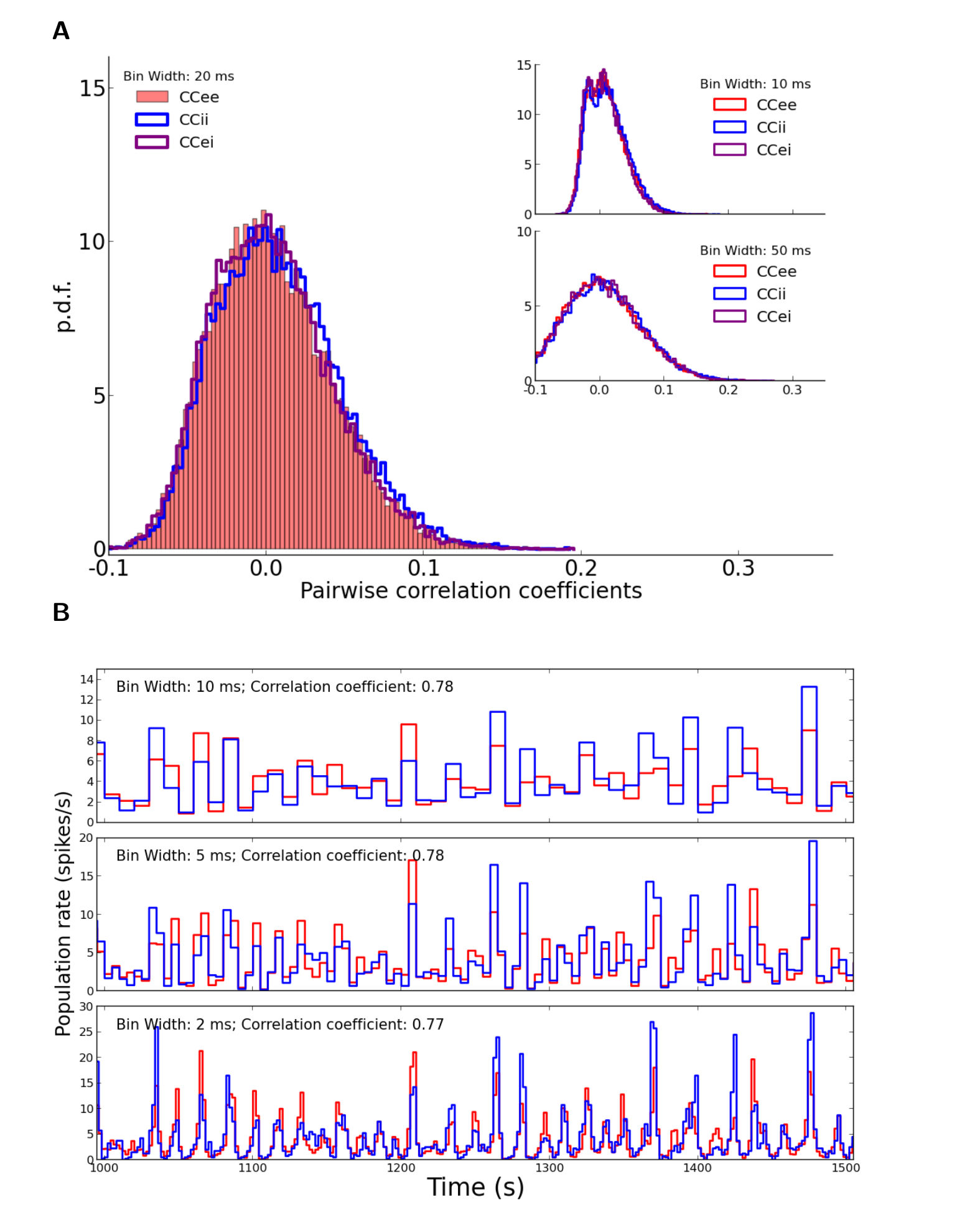} 

\caption{{\bf Correlations in a network with distance-dependent connectivity.} 
Distribution of correlation coefficients for pairs of neurons \textbf{(A)} and temporal correlation of population activities \textbf{(B)} in the example network of Fig.~\ref{Fig_DistDepCon} with distance-dependent connectivity. 
Other conventions are similar to Fig.~\ref{Fig_ER_Corr}. 
} 
\label{Fig_CCNLG} 
\end{figure}

In fact, however, pairwise correlations do not seem to be systematically larger than in random networks Fig.~\ref{Fig_ER_Corr}A, judged by the distribution of Pearson correlation coefficients (Fig.~\ref{Fig_CCNLG}A). 
In contrast, the fluctuations in the activity of excitatory and inhibitory populations seem to be even less correlated (compare Fig.~\ref{Fig_CCNLG}B with Fig.~\ref{Fig_ER_Corr}B). 
Occasional partial imbalance of excitatory and inhibitory input may therefore cause systematic distortions of our linear prediction. 



\begin{figure}   

\centering\includegraphics[width=6.0in]{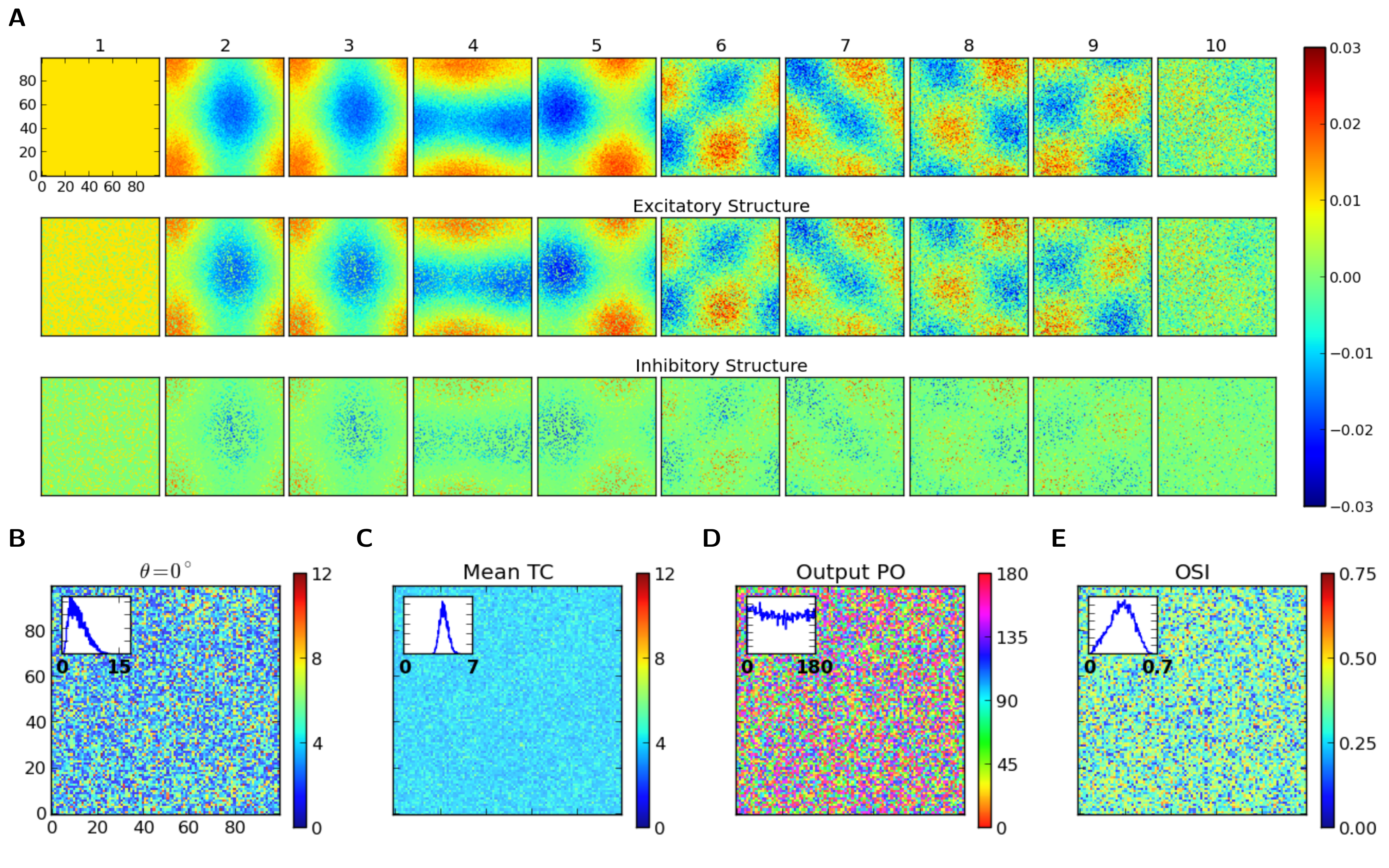} 

\caption{{\bf Structure and dynamics of a network with distance-dependent connectivity.} 
\textbf{(A)}~First ten eigenvectors, corresponding to the ten eigenvalues of largest magnitude, are plotted for the sample network described and discussed in Figs.~\ref{Fig_DistDepCon} and \ref{Fig_DDC}. 
For each eigenvector, the value of the vector corresponding to each neuron is plotted at the respective spatial position of the neuron (as in Fig.~\ref{Fig_DistDepCon}A).
In the first row, this is shown for all neurons, and in the bottom rows, the structure of eigenvectors are separately plotted for excitatory and inhibitory neurons, respectively (with zeros replaced on the positions of the other population, respectively).
Only the real part of the components of the eigenvectors are plotted here. 
Note that the tenth eigenvector already corresponds to an eigenvalue from the bulk of the spectrum in Fig.~\ref{Fig_DistDepCon}D. 
\textbf{(B)}~Shown is the mean firing rate of neurons in the network, extracted from a $15\,\mathrm{s}$ simulation, in response to a stimulus with orientation $\theta = 0\deg$. 
\textbf{(C)}~For each neuron, the mean tuning curve (Mean TC) is plotted as the average (over different orientations) of the mean firing rate. 
\textbf{(C, D)}~From each tuning curve, $r(\theta)$, the output preferred orientation (Output PO) and output orientation selectivity index (Output OSI) is extracted and plotted, respectively. 
They are obtained as the angle and length of the orientation selectivity vector, $\mathrm{OSV} = \sum_\theta r(\theta)\exp(2\pi i\theta/180\deg) / \sum_\theta r(\theta)$; i.e.\ $\mathrm{OSI} = |\mathrm{OSV}|$ and $\mathrm{PO} = \arg(\mathrm{OSV})$. 
Insets show the distributions in each case. 
} 
\label{Fig_EV} 
\end{figure}

Another potential contributor to the discrepancy of predictions are the different structural properties of these networks, reflected among other things in their respective eigenvalue spectrum.  
It is therefore informative to look more carefully into the eigenvalues which mark the difference to Erd\H{o}s-R\'{e}nyi networks, i.e.\ the ones localized between the bulk spectrum and the exceptional eigenvalue corresponding to the common-mode. 
To evaluate this, the first ten eigenvectors (corresponding to the ten largest eigenvalues sorted by their magnitude) of the network are plotted (Fig.~\ref{Fig_EV}A). 
The first eigenvector is the uniform vector (common-mode), and the tenth one is hardly distinguishable from noise. (Note that the corresponding eigenvalue is already part of the bulk.) 
In between, there are eight eigenvectors with non-random spatial structure. 

These eigenvectors reflect the specific sample from the network ensemble we are considering here, and they can, in principle, prefer a specific pattern of stimulation in the input. 
While other patterns of input stimulation would be processed by the network $W$ with a small gain, any input pattern matching these special eigenmodes would experience the highest gain (in absolute terms) from the network. 
The corresponding eigenvalues $\lambda$ have, however, a negative real part, therefore these modes would in this case be attenuated: 
the corresponding eigenvalues of the operator $A = (\one -W)^{-1}$ that yields the stationary firing rate vector, namely $\lambda_A = \frac{1}{1-\lambda_W}$, would then be very small. 

We do not, however, explicitly represent any of these patterns in our stimuli. 
The stimuli considered in this work can be broken down to a linear sum of the common-mode (i.e.\ the first eigenvector) and the modulation component (i.e.\ a random pattern, as preferred orientations are assigned randomly and independently to all neurons, irrespective of the position of the neuron in space). 
The modulation component would therefore only have a very small component in the direction of each any special eigenmode. 
It is however possible that for non-stationary inputs to the network, transient patterns with a bias for selected eigenmodes resonate more than others. 

The question arises, if spatially structured eigenmodes (cf.~Fig.~\ref{Fig_EV}A) have an impact on the observed pattern of spontaneous and evoked neuronal activity. 
Plotting the response of the network to a stimulus reflecting one particular orientation, as well as the mean activity of neurons over different orientations, do not reveal any visible structure (Fig.~\ref{Fig_EV}B, C). 
The baseline activity of the network seems to be quite uniform, and the response to a certain orientation does not reveal any structure beyond the random spatial pattern one would expect from the random assignment of preferred orientations of the input. 
This is further supported by visual inspection of the map of preferred orientations for the output (Fig.~\ref{Fig_EV}D) and orientation selectivity index (Fig.~\ref{Fig_EV}E) in the network. 

In principle, it is conceivable that  spatially structured eigenmodes could affect the response of the network by setting the operating point of the network differently at different positions in space, as a result of the selective attenuation of certain eigenmodes. However, we have never observed such phenomena in our simulations. 
The fact that those structured modes get attenuated (and not amplified) might be one reason; another reason might be the fact that eigenmodes are typically heterogeneous and non-local, which makes the selection of the corresponding overall preferred pattern unlikely. 
Spatial structure of the network, and of its built-in linear eigenmodes, are therefore not dominant in determining the distribution of orientation selectivity. 
They could, however, be potential contributors in the small deviation of the predicted distribution from the measured one. 



\subsection*{Spatial Imbalance of Excitation and Inhibition} 

\begin{figure}   

\centering\includegraphics[width=6.0in]{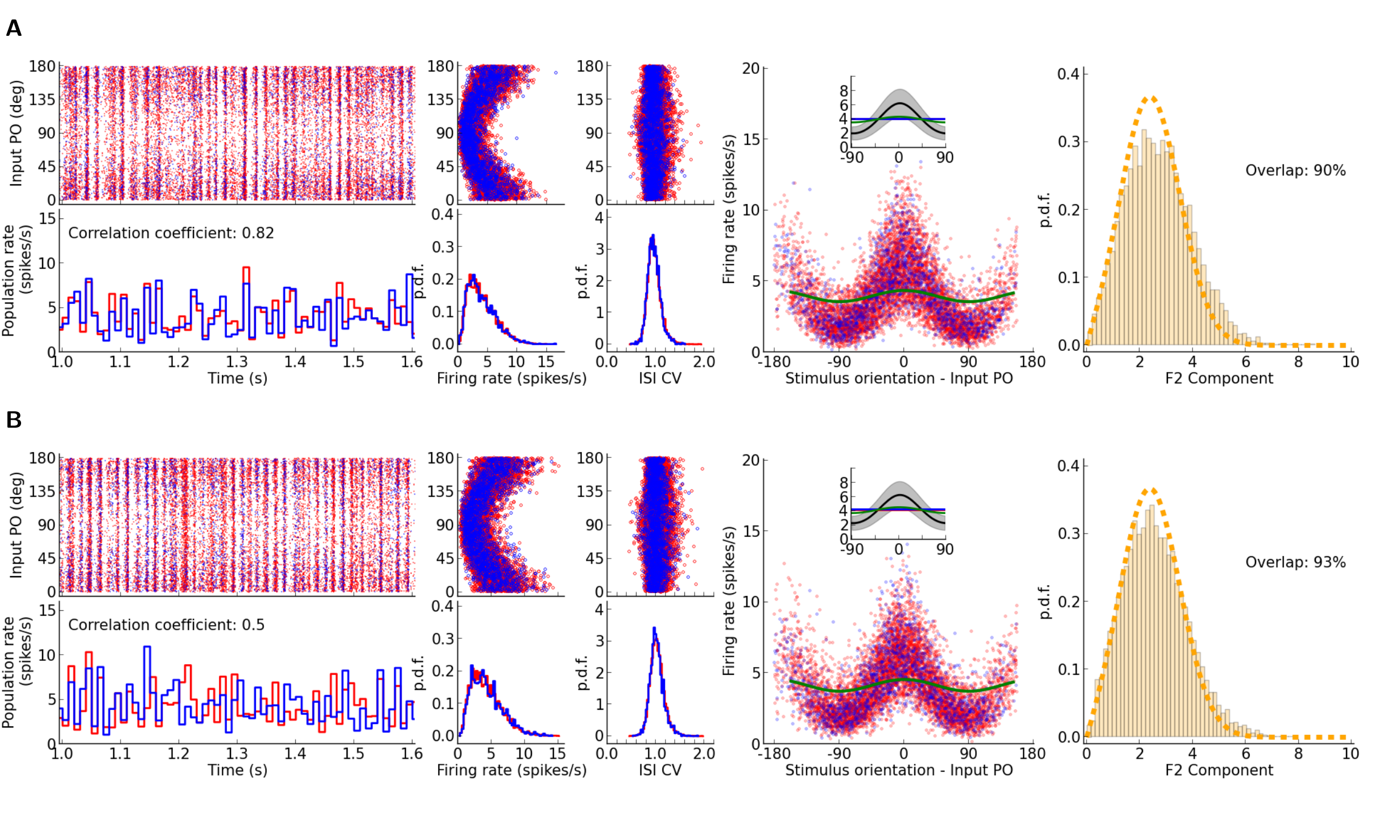} 

\caption{{\bf The impact of spatial extent of excitation and inhibition on the distribution of F2 components.} 
Same illustration as in Fig.~\ref{Fig_ER_Rec}, for simulations with different extents of excitatory and inhibitory connectivity. 
\textbf{(A)} shows the results for a network with inhibition being more localized than excitation ($\sigma_\mathrm{inh} = 0.45\,\mathrm{mm}$ and $\sigma_\mathrm{exc} = 0.75\,\mathrm{mm}$). In \textbf{(B)} we show the results for excitation being more localized than inhibition ($\sigma_\mathrm{inh} = 0.75\,\mathrm{mm}$ and $\sigma_\mathrm{exc} = 0.45\,\mathrm{mm}$). 
Other parameters are the same as in Fig.~\ref{Fig_DDC}. 
The distribution of F2 components is computed after re-normalizaton of the connectivity matrix by $\zeta_s$, as explained before. 
} 
\label{Fig_SpatDisb} 
\end{figure}

To test the robustness of our predictions, we went beyond the case of spatial balance of excitation and inhibition, and also simulated networks with different extents of connectivity. 
Roughly the same overall behavior of the network, and accuracy of our predictions, were observed for the case of more localized inhibition and less localized excitation ($\sigma_\mathrm{inh} = 0.45\,\mathrm{mm}$ and $\sigma_\mathrm{exc} = 0.75\,\mathrm{mm}$, Fig.~\ref{Fig_SpatDisb}A), as well as for the case of more localized excitation and less localized inhibition ($\sigma_\mathrm{inh} = 0.75\,\mathrm{mm}$ and $\sigma_\mathrm{exc} = 0.45\,\mathrm{mm}$, Fig.~\ref{Fig_SpatDisb}B). 



\begin{figure}  

\centering\includegraphics[width=6.0in]{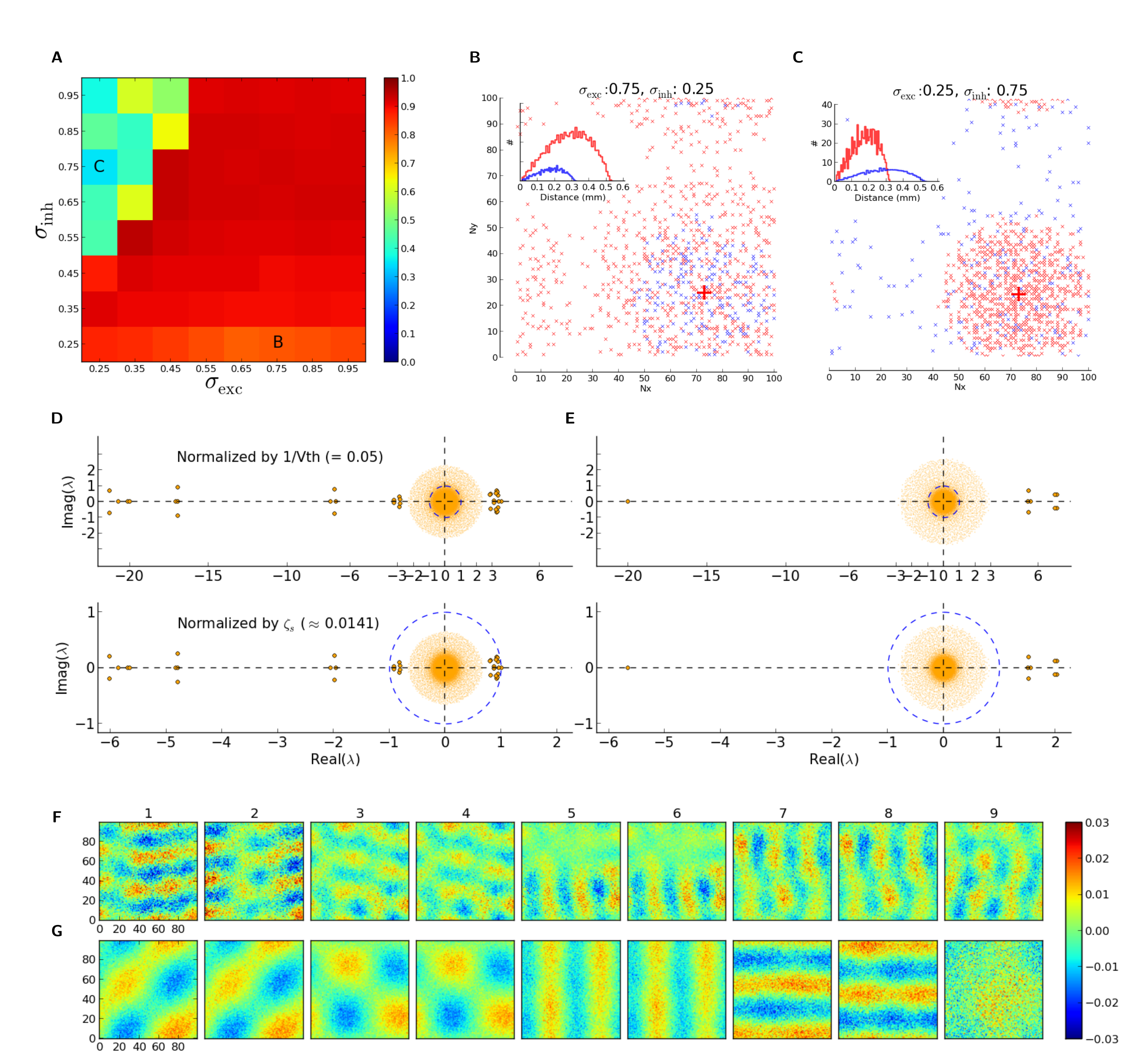} 

\caption{{\bf Accuracy of the linear prediction for different spatial extents of excitation and inhibition.} 
\textbf{(A)}~The overlap index (using $\zeta_s$) is plotted for networks with different extents of excitation and inhibition. 
\textbf{(B, C)}~Pre-synaptic connections for a sample post-synaptic neuron, along with the histogram of distances to pre-synaptic neurons for the entire population (inset), are shown here for two extreme cases, marked in panel (A). 
\textbf{(D, E)}~Eigenvalue distribution of the example networks in (B) and (C), respectively. 
Two ways of normalization of the weight matrix are compared in the top and bottom panels. 
\textbf{(F, G)}~First nine eigenmodes, corresponding to the nine largest positive eigenvalues (in terms of their real component), are plotted for the example networks in (A). 
Panels (F) and (G) correspond to the networks in (B) and (C), respectively. 
Note that the ninth eigenvector in (G) corresponds to an eigenvalue from the bulk of the spectrum in (E). 
Only the real part of the components of the eigenmodes are plotted. 
} 
\label{Fig_DiffExt} 
\end{figure}

This trend was further corroborated when we systematically scanned the accuracy of our predictions for a large set of different networks, by scanning the parameter space (Fig.~\ref{Fig_DiffExt}A). 
Indeed, for most of the parameters studied, the predicted distribution of orientation selectivity matched very well with the actual distribution (more than $90\%$ overlap).  
For the more ``extreme'' combinations of parameters, however, where the spatial extent of excitation and inhibition were highly out of balance, the quality of the match degraded. 
The deviation was more significant when excitation was more local and inhibition was more global (Fig.~\ref{Fig_DiffExt}A, upper left portion). 
Note that, even for the most extreme cases of local excitation ($\sigma_\mathrm{exc} = 0.25\,\mathrm{mm}$), the accuracy of our prediction is still fairly good, as long as the inhibition has a similar extent ($\sigma_\mathrm{inh} = 0.25$ -- $0.45\,\mathrm{mm}$). 

To investigate what happens in each extreme case, we chose two examples (marked in Fig.~\ref{Fig_DiffExt}A) for further analysis. 
The connectivity patterns of these two examples, with $(\sigma_\mathrm{exc}, \sigma_\mathrm{inh}) = (0.75, 0.25)$ and $(0.25, 0.75)$ (numbers indicated in $\mathrm{mm}$), are illustrated in Fig.~\ref{Fig_DiffExt}B, C, respectively. 
The eigenvalue spectra of the corresponding weight matrices are shown in Fig.~\ref{Fig_DiffExt}D, E. 
When the weights are normalized with respect to the reset voltage (upper panels), both spectra suggest an unstable linearized dynamics, as they both have eigenvalues with a real part larger than one. 

The picture changes, however, when a normalization according to the effective gain, $\zeta_s$, is performed. 
While the network with local excitation still has several clearly unstable eigenmodes (Fig.~\ref{Fig_DiffExt}E, bottom), the spectrum of the network with local inhibition comprises only one positive eigenvalue which is only slightly larger than one (Fig.~\ref{Fig_DiffExt}D, bottom). 
Some of the eigenvectors corresponding to the largest positive eigenvalues are plotted for both networks in Fig.~\ref{Fig_DiffExt}F, G, respectively. 
From this, it seems therefore possible that the source of deviation from the linear prediction is indeed instability of the linearized dynamics (namely the instability of the uniform asynchronous-irregular state about which we perform the linearization) for these extreme parameter settings. 
When this instability is more pronounced, i.e.\ for the network with local excitation, the deviation is highest. 
When the network is at the edge of instability, i.e.\ for the network with local inhibition, our predictions show only a modest deviation. 



\begin{figure} 

\centering\includegraphics[width=6.0in]{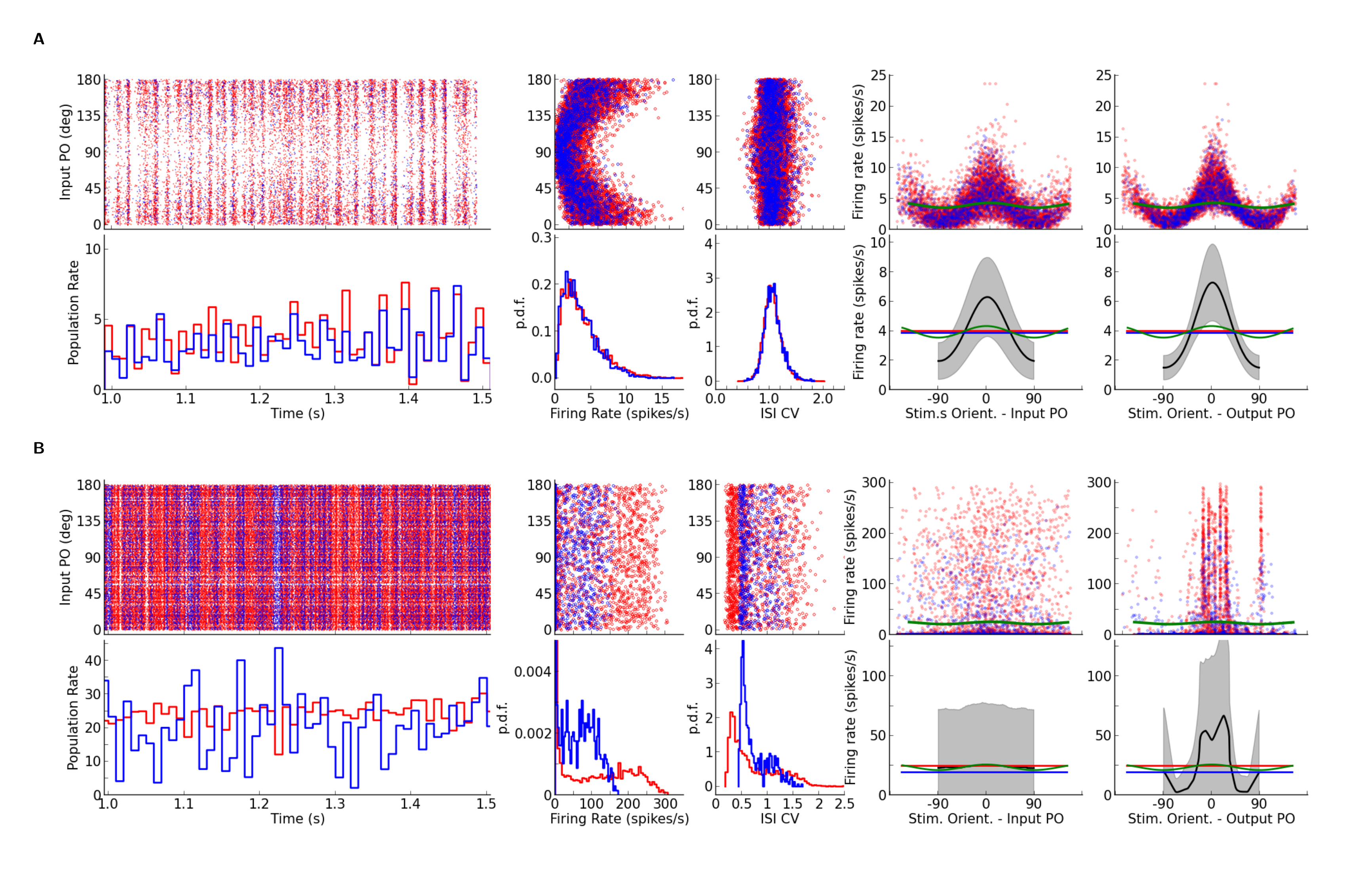} 

\caption{{\bf Orientation selectivity in networks with extreme spatial imbalance of excitation and inhibition.} 
\textbf{(A, B)}~As extreme examples, networks with highly local inhibition ($\sigma_\mathrm{inh} = 0.25\,\mathrm{mm}$ and $\sigma_\mathrm{exc} = 0.75\,\mathrm{mm}$, Fig.~\ref{Fig_DiffExt}B) or highly local excitation ($\sigma_\mathrm{inh} = 0.75\,\mathrm{mm}$ and $\sigma_\mathrm{exc} = 0.25\,\mathrm{mm}$ , Fig.~\ref{Fig_DiffExt}C), were considered, respectively. The spiking activity of the network (first column), distribution of firing rates (second column) and spike train irregularity index (third column), as well as output tuning curves (fourth and fifth columns). 
In the fourth column, the tuning curves are aligned according to their Input PO, whereas in the fifth column they are aligned according to their Output PO. 
Other conventions are the same as Fig.~\ref{Fig_SpatDisb}. 
} 
\label{Fig_Extremes} 
\end{figure}

To test this hypothesis further, namely that instability of the linearized dynamics is the source of mismatch between the linear prediction and the actual distribution of orientation selectivity, we need to scrutinize the response behavior of the sample networks. 
The outcome of this is shown in Fig.~\ref{Fig_Extremes}. 
While the network with local inhibition does not look very different from other examples considered before (Fig.~\ref{Fig_Extremes}A), the behavior of the network with local excitation very clearly shows deviating behavior (Fig.~\ref{Fig_Extremes}B). 
First, firing rates are much higher than in the less extreme cases, for both excitatory and inhibitory populations (Fig.~\ref{Fig_Extremes}B, first column). 
Moreover, the activity of excitatory and inhibitory neuronal populations are not well correlated in time, as it is the case for the other networks (Fig.~\ref{Fig_Extremes}B, first column, bottom). 
The firing rate distribution has a very long tail, and the tail is longer for the excitatory than for the inhibitory population (Fig.~\ref{Fig_Extremes}B, second column). 
The long tail is accompanied by a peculiar peak at zero firing rate (which is cut for illustration purposes in Fig.~\ref{Fig_Extremes}B, second column, bottom). It reflects the fact that most of the neurons in the network are actually silent, and a small fraction of the population is highly active. 
The average irregularity of spike trains (the CV of the inter-spike intervals) in the network is reduced compared to our previous examples (Fig.~\ref{Fig_Extremes}B, third column). 
All these properties are consistent with the presumed instability of the linearized dynamics, as inferred from the eigenvalue spectrum. 

In terms of functional properties of the network, the output tuning curves are much more scattered when aligned by the respective preferred orientations of the inputs (Fig.~\ref{Fig_Extremes}B, fourth column, upper panel). 
In fact, the mean output tuning curve for all neurons of the network does not show any amplification, if it is aligned at the Input PO (Fig.~\ref{Fig_Extremes}B, fourth column, lower panel). 
The picture changes, however, if tuning curves are aligned according to their Output PO (Fig.~\ref{Fig_Extremes}B, fifth column). 
Here a clear amplification of the modulation is evident in output tuning curves, although the relation to the feedforward input gets lost. 
Also, the average output tuning curve is not smooth, i.e.\ not all orientations are uniformly represented in the distribution of output preferred orientations. 



\begin{figure}   

\centering\includegraphics[width=6.0in]{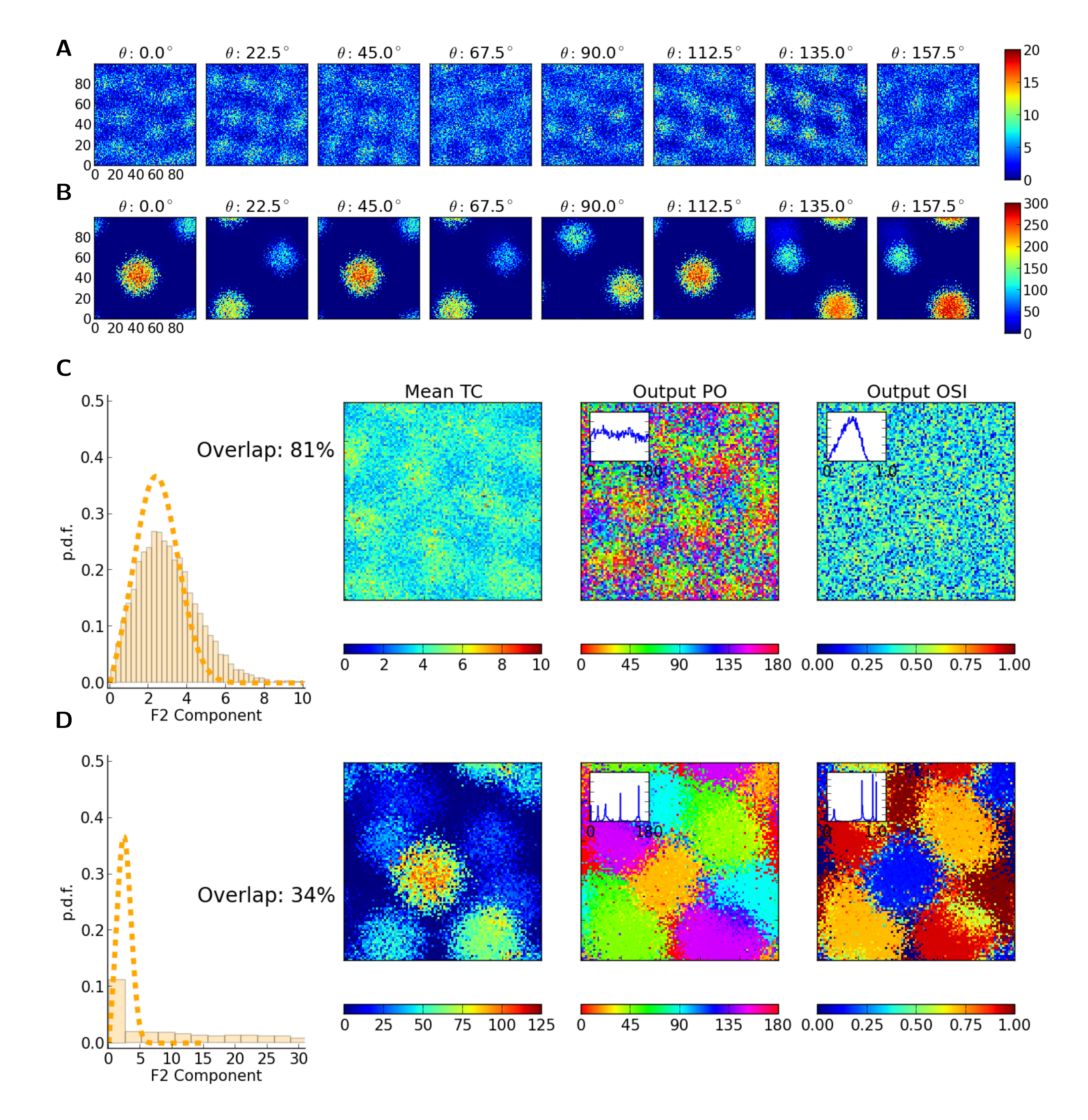} 

\caption{{\bf Dynamic instability leads to nonlinear distortions in the processing of orientation selectivity.} 
\textbf{(A, B)}~Mean firing rate of neurons in the network (the same networks as in Fig.~\ref{Fig_Extremes}, (A) and (B), respectively) in response to stimuli of different orientations. 
\textbf{(C, D)}~For the networks in (A) and (B), respectively, the distributions of F2 components are compared with the linear prediction (using $\zeta_s$, dashed line) in the first column. 
The subsequent columns depict the map of average (over orientation) tuning curves (Mean TC), Output PO and Output OSI. Insets in the last two columns show the distribution of Output PO and Output OSI, respectively. 
} 
\label{Fig_Instab} 
\end{figure}

This breaking of the symmetry becomes even more obvious when we look at the response of the two networks to stimuli of different orientations (Fig.~\ref{Fig_Instab}A, B). 
While both networks show some degree of inhomogeneity in the spatial pattern of their firing rate responses, the response pattern of the second network is much more clustered (Fig.~\ref{Fig_Instab}B). 
In fact, it seems that the internal connectivity structure of the network determines the position of a discrete set of potential activity bumps, and the orientation bias in the input can only choose between these bumps. 
As the nonlinear dynamics of the unstable network is crucially affecting the activity in response to stimuli, it is not surprising that the distribution of orientation selectivity is not matching the prediction which relies on a linearization about the uniform asynchronous-irregular state (compare Fig.~\ref{Fig_Instab}C and D, first columns). 

In fact, this internal structure is even reflected in the pattern of baseline firing rates (mean of the tuning curves over orientation). 
While for the network with local inhibition this pattern is covert and ineffective (Fig.~\ref{Fig_Instab}C, second column), in the network with local excitation clear clusters of activity, resembling the ones in Fig.~\ref{Fig_Instab}B, are evident (Fig.~\ref{Fig_Instab}D, second column). 
One may, therefore, expect that there exists a corresponding pattern in the spatial organization of orientation selectivity. 
Larger domains of neighboring neurons, who get activated together, also exhibit the same selectivity. 
This is reflected in the clustering of output preferred orientations (Fig.~\ref{Fig_Instab}C, third column) and orientation selectivity index (Fig.~\ref{Fig_Instab}C, fourth column). 

Note that a consequence of this clustering of PO is a degenerate representation of orientation selectivity, i.e.\ not all orientations are represented equally in the network. 
While the distribution of Output POs is almost uniform in the network with local inhibition (inset in Fig.~\ref{Fig_Instab}C, third column), clear peaks are present in the distribution of Output POs in the network with local excitation (inset in Fig.~\ref{Fig_Instab}D, third column). 
This is in line with our observation of broken symmetry described before, reflected in the pattern of mean output tuning curve in Fig.~\ref{Fig_Extremes}B. 



\section*{Discussion} 

We presented a linear analysis, which was capable of predicting the distribution of orientation selectivity in networks with different patterns of random connectivity, including some degree of spatial organization, and for a wide range of parameters. 
The effective strength of excitation and inhibition in the network (Figs.~\ref{Fig_ER} and Fig.~\ref{Fig_ER_Rec}), as well as the spatial extent of excitatory and inhibitory connectivity (Fig.~\ref{Fig_DiffExt}), did not affect the prediction accuracy very strongly, as long as the linearized dynamics remained stable. 
We therefore conclude that linear mechanisms are the major network operations that explain amplification and attenuation, and the distribution of the resulting orientation selectivity in our networks, within their stable regimes of linearized dynamics. 
 
\subsection*{Operating Regime of Orientation Selectivity} 
 
Note that even in networks with localized connectivity of excitation and/or inhibition, the linearized dynamics remained stable for a vast set of parameter combinations. 
Even when excitation was highly local and clustered, as long as inhibition had the same spatial connectivity profile, stability of the network was guaranteed. 
A similar conclusion has been recently obtained from an analysis of spatially embedded balanced networks \cite{Rosenbaum2014}. 
It has also been shown before that networks with distance-dependent connectivity can show the same macroscopic behavior similar to random networks without local connectivity \cite{Yger2011}. 

The asynchronous irregular (AI) state has been argued to best match the activity of cortical networks \textit{in vivo} (see e.g.\ \cite{vanVreeswijk1996, vanVreeswijk1998, Amit1997}). 
The relevance of this regime has only been discussed, however, for cortical networks in response to uniform stimulation.
On the other hand, with regard to the processing of a non-uniformly modulated input, it has been claimed that a ``marginal state of recurrent dynamics'' might be the relevant regime of operation for the processing of weakly tuned inputs \cite{Ben-Yishai1995}. 
Also, it has recently been suggested that a recurrent regime with ``macroscopic chaos'' (probably corresponding to our regime of unstable dynamics) might be advantageous for sensory processing, as it may support a better separation of trajectories \cite{Ostojic2014}.

In contrast to these proposals, the results of our study suggest that a stable AI state of of dynamics might indeed be the relevant regime of operation also for sensory processing in cortical networks in response to tuned inputs. 
Notably, the dense and local pattern of inhibition in real cortical circuits \cite{Fino2011, Packer2011, Hofer2011} is in line and consistent with our proposal. 
It might indeed be a general strategy biological networks of spiking neurons have exploited to ensure their overall stability to modulated inputs. 
We note again that we are talking about dynamic stability here, where the network dynamics is linearized about the uniform asynchronous-irregular state, and the effective weights of coupling linearized about this baseline state are considered. 
 
\subsection*{Distribution of Orientation Selectivity} 

A broad distribution of orientation selectivity is reported across all cortical layers in the primary visual cortex of macaque monkeys \cite{Ringach2002}, as well as in mice \cite{Niell2008} (for a comparison of the distributions, see panel C in Fig.~S2 therein).
Although we chose random connectivities by fixing the in-degree of all neurons (which we refer to as ``structural homogeneity''), a broad distribution of orientation selectivities also emerged in all our networks. The main contributor to this broad distribution was, therefore, not the structural heterogeneity of synaptic connectivity. In fact, there is no heterogeneity at all, if one only considers the number of connections each neuron receives from pre-synaptic excitatory and inhibitory neurons
Nor were the temporal fluctuations of activity generated by our networks a major source of this variability, although the networks were mostly operated in the fluctuation-driven regime with high amounts of temporal and trial-by-trial variability. 
As we have generally chosen a homogeneous connectivity pattern, this temporal variance would be essentially the same for all neurons, at least in the baseline state. (This also justifies the mean-field ansatz we have employed for our analysis.) 
This is again reflected in the narrow distribution of F0 components in all our networks. 

The main source of variability in orientation selectivity is rather the ``functional heterogeneity'' in synaptic connectivity, 
namely heterogeneous preferred features (here, preferred orientations of inputs) of the pre-synaptic sources within the recurrent network. 
Receiving input from neurons with different preferred features may be a computational strategy to integrate the information, and help to remove distractive correlations in the activity. 
The fact that each neuron within the recurrent network receives input from a heterogeneous pool of neurons with a wide range of preferred orientations leads to a random ``summation'' of pre-synaptic preferred orientations, which eventually changes the output preferred orientation of the post-synaptic neuron \cite{Sadeh2014}. 

The quenched noise of preferred orientations, and not structural or dynamic fluctuations, is, therefore, the main mechanism responsible for the distribution of orientation selectivity in our networks. 
We showed that even with this most conservative estimate of neuronal heterogeneity, consistent with recent experiments \cite{Jia2010}, a broad distribution of neuronal selectivities can be obtained. 
However, we cannot rule out a possible contribution of other sources of heterogeneity, like heterogeneous connectivity and heterogeneous amounts of excitation and inhibition different neurons may receive in their baseline state (leading to different levels of spontaneous activity, see e.g.\ \cite{Ringach2002}), as well as variability in neuron parameters \cite{Yim2013} and synaptic noise. 
Also, heterogeneity in the pattern of feedforward projections to neurons in V1 can be a prominent source of distribution in orientation selectivity.
However, if the distribution of orientation selectivity is mainly dominated by feedforward heterogeneity, or / and if single neuron heterogeneities like variability in threshold and synaptic noise are the main source of this distribution, the distribution should not much change when the recurrent network is absent.
On the other hand, if functional heterogeneity resulting from recurrent interactions is a major contributor to this distribution, it should get narrowed when the intra-cortical circuitry is deactivated.
It therefore awaits further experimental tests which mechanisms are dominant in creating the distribution of feature-selectivity in the cortex.

\subsection*{Future Directions} 

There are several ways in which the the current study could be expanded. 
First, sticking to a linear framework of analysis enabled us to analytically compute the distribution of orientation selectivity. 
In this simplified framework, however, we neglected several nonlinearities, both at the level of neuronal properties and network interactions. 
These nonlinearities are deemed to be more prominent in biological networks, for instance in the form of rectification \cite{Carandini2000, Priebe2008}, or an expansive-compressive transfer nonlinearity \cite{Anderson2000, Miller2002, Priebe2012}. 
Such mechanisms might play a major role in sharpening and amplification of orientation selectivity. 
A more complete theoretical treatment of the problem should therefore consider the contribution of nonlinear mechanisms as well, although this may come at the expense of less rigorous analytical predictions. 

One way to embrace additional nonlinear mechanisms that are effective in biological networks, at least at the level of simulations, is to use a more realistic and more detailed neuron model. 
In our simulations here we used the current-based LIF neuron model. 
Simulating networks of more realistic neuron models, like conductance-based LIF neurons, may change certain behaviors of the network \cite{Kuhn2004, Kumar2008}. 
For instance, increasing the recurrent coupling in our inhibition-dominated networks can decrease the mean membrane potential of neurons in the network to very negative values, as there is no reversal potential limiting it. 
This is not the case in a conductance-based neuron model, and therefore a network of that sort might show a different behavior, especially when operated in extreme regimes. 
 
Finally, it would be interesting to see how the predictions of our current theory change when one considers networks with feature-specific connectivity. 
This scenario might be corresponding to species with orientation maps, where neighboring neurons tend to have a similar preferred orientation \cite{Blasdel1986, Bonhoeffer1991, Ohki2006}, or to species without spatial map of selectivity, but with feature-specific functional connectivity \cite{Hofer2011, Ko2011, ko2013, Ko2014, Ishikawa2014}. 
A linear amplification of feedforward input, for instance, has been recently reported in cortical circuits of mice \cite{Lien2013, Li2013, Li2013a}. 
How this effect could be modeled within our theoretical framework, and how it affects the distribution of orientation selectivity, should therefore be a next step in our research. 



\section*{Acknowledgments} 

We thank the developers of the simulation software NEST (see http://www.nest-initiative.org) and the maintainers of the BCF computing facilities for their support throughout this study. 
Funding by the German Ministry of Education and Research (BMBF; BFNT-F*T, grant 01GQ0830) and the German Research Foundation (DFG; grant EXC 1086) is gratefully acknowledged. The article processing charge was covered by the open access publication fund of the University of Freiburg.




\clearpage{}


\end{document}